\documentclass[default,iicol]{sn-jnl}% Default with double column layout

\jyear{2021}%

\theoremstyle{thmstyleone}%
\theoremstyle{thmstyletwo}%

\theoremstyle{thmstylethree}%

\begin{document}

\title[Article Title]{Searching for initial state fluctuations in heavy ion collisions at FAIR energy using Principal Component Analysis}


\author*[]{\fnm{Ekata} \sur{Nandy}} \email{ekata@vecc.gov.in}

\author[]{\fnm{Subhasis} \sur{Chattopadhyay}} \email{sub.chattopadhyay@gmail.com}


\affil*[]{\orgdiv{Variable Energy Cyclotron Centre},  \orgaddress{\street{1/AF,Bidhan Nagar}, \city{Kolkata}, \postcode{700064}, \state{WB}, \country{India}}}

\iffalse
\affil[2]{\orgdiv{Department}, \orgname{Organization}, \orgaddress{\street{Street}, \city{City}, \postcode{10587}, \state{State}, \country{Country}}}

\affil[3]{\orgdiv{Department}, \orgname{Organization}, \orgaddress{\street{Street}, \city{City}, \postcode{610101}, \state{State}, \country{Country}}}
\fi

\abstract{In high energy heavy ion collisions, the initial configurations of the colliding nuclei play an important role in determining the reaction type and the products of the reaction. The initial arrangement of nucleons within the overlap region of two colliding nuclei is generally asymmetric and such asymmetries reflect themselves in the measurement final state momentum anisotropy. Also initial distribution of the nucleons are subjected to large quantum fluctuation causing large energy deposition in a small region. The final state observables related momentum anisotropies although sensitive to such localized fluctuations but their true effect gets diluted because these observables are calculated by averaging over a set of events. Also, such fluctuations in the initial states are random and uncontrolled. Thus, identifying their effect from event-averaged final state observable is difficult. However, it would be interesting to know the origin of such fluctuations and how these fluctuation are eventually translated to the final state. In this work, we at first introduce such localized fluctuations in the initial configurations, also called hot spots, by implementing spatial rearrangements of nucleon position in the colliding nuclei in the central Pb+Pb collisions at E$_{lab}$=20 AGeV ($\sqrt{s}$=6.27 GeV) using the UrQMD event generator. Then the final state distributions of one or two dimensional variables e.g., ($\eta$, $\phi$, $p_T$) and ($\eta-p_T$, $\phi-p_T$, $\eta-\phi$) of the produced pions are analysed using the principal component analysis (PCA) technique. The eigenvalues of the principal components have been studied for various initial configurations, event fractions containing hot spots in the initial condition and for event centralities with an aim to find it's sensitivity to the initial hot spot configurations.}




\maketitle

\section{Introduction}\label{sec1}
In collisions of heavy ions at high energy, the initial distributions of the colliding nucleons or of the partons play important roles in deciding the subsequent stages of the formation and evolution of the medium and of the observables that are based on the final state particles \cite{star,phenix}. The signatures of the type of the medium formed could only be understood if we can model or understand the initial state properly. Depending on the energy of the collisions, various types of models like Glauber model \cite{glauber,glauber2}, colour glass condensate (CGC) \cite{CGC} are frequently being used to describe the initial states of the collisions. Conclusions about the correct descriptions of the initial state configurations are made based on the model that describes the data well \cite{CGC_data}. However, there is no established unique method which could be applied to probe solely the initial state of the collisions. It has been shown that the structures in the initial nucleonic or partonic states might be carried over by some observables that remain unaffected by the evolution of the medium. One type of such observables is the flow coefficients of various orders that reflect the momentum anisotropy in the final state particles as representatives of the initial state spatial anisotropies \cite{flow1, flow2, flow3}. Extensive studies have been done in this area in heavy ion collisions spanning over a range of colliding energies and collision species. The prominent feature of the initial state configurations in such cases is that the arrangements of the colliding nucleons over a large number of events could be described as the convolution of several geometrical shapes like ellipse, triangle etc \cite{flow3_data}. These geometries appear where the models like Monte-Carlo Glauber \cite{mc_glauber} with Woods-Saxon potential is used to describe the geometry of the collision. These asymmetries are studied experimentally with the help of the Fourier decomposition of the azimuthal distributions with respect to the reaction plane \cite{flow4}. By now, it is well known that the initial conditions in heavy ion collisions are not smooth rather, it fluctuates from event to event. This event-by-event fluctuation of the collision geometry is a result of random fluctuations in the nucleon positions within the overlap region of two colliding nuclei. Such fluctuations may sometime cause large energy density in the localized regions known as the hot spots~\cite{hotspot,hotspot2,hotspot3}.
Following the collision, as the medium evolves, these hot spots are also translated to the final state causing large fluctuations in local domains. But, experimental measurements have not been able to capture such local fluctuations till date. In this regard, there are some open questions that need to be addressed; what is the spatial extent of these hot spots ? and whether the fluctuations in the final state depend on the initial size of the hot spots?

In this work we have explored the sensitivity to these event wise initial state fluctuations by adopting a generalized approach where the phase-space momentum distributions of the final state particles  are decomposed by using the Principal Component Analysis (PCA) technique \cite{pca_1, pca_2}.
This technique reduces the dimension of the correlated space and the eigenvalues of the prominent components represent the variance along that component. As the initial fluctuations could be imprinted in the distributions of any particular or in all final state variables, for a systematic study, we first implemented a spatial rearrangement of the nucleons on top of the Woods-Saxon distribution so that hot spot like local regions could be created in the initial state. Subsequently, we study the PCA eigenvalues of the final state particles under various initial conditions as inputs.
We have worked at an intermediate energy E$_{lab}$=20 AGeV ($\sqrt{s}$=6.27 GeV) Pb+Pb collisions, which is likely to be accessible at the beam energy scan programme of RHIC \cite{bes-rhic} or at the upcoming FAIR facility \cite{fair} and at this energy, the initial state could be well described using nucleons and therefore we have introduced the hot spots like initial configurations at the nucleonic level. We have used UrQMD \cite{urqmd} as the model for the transportation of the colliding nuclei, evolution of the medium and the production of the final state particles. At this energy, UrQMD describes most of the available data quite well.
The article is organised as follows, the UrQMD model and its applicability at this energy has been discussed in the next section. We then discussed PCA and its implementation in the present study in the section 3. The implementation of hot spots in the initial configuration has been discussed in the section 4, followed by results in section 5 and summary in the section 6.

\section{The UrQMD Model}

UrQMD (Ultra-relativistic Quantum Molecular Dynamics) is a non-equilibrium microscopic transport model that solves the relativistic Boltzmann transport equation
%\vspace{-3.0}
\begin{equation}
\textit{p}^{\mu}\partial_{\mu}\textit{f}_{i}(\textit{x}^\nu,\textit{p}^\nu) = \textit{C}_{i}
\end{equation} 
%\vspace{-2.0}
This describes the time evolution of the distribution function, $\textit{f}_{i}$ in coordinate and momentum space for a particle species "$\textit{i}$"  and includes the full collision term $\textit{C}_{i}$. Full particle-antiparticle, isospin and flavor SU(3) symmetries are applied. An individual particle propagates on a straight line until the relative distance between two particles is smaller than a critical distance given by the total interaction cross-section between two particles. These cross-sections in UrQMD are either calculated by the principle of detailed balance or additive quark model \cite{aqm1,aqm2} or parametrized from the available experimental data. For resonance excitations or decays the Breit-Wigner formalism is used. Until now, UrQMD has 55 Baryons and 32 Mesons that include ground state particles and all resonances with mass upto 2.25 GeV. 
UrQMD model is a very successful model in describing the dynamics of particle production from p+p, p+A to A+A collisions over a broad energy range \cite{urqmd-hh,kalyan}. Below the collision energies of, $\sqrt{s_{NN}} < $ 5 GeV the dynamics of particle production is governed by hadronic interactions. Whereas, at higher energies, a concept of excited color string formation followed by hadronization via string fragmentation is invoked to introduce the effect of partonic degrees of freedom \cite{urqmd}.

In UrQMD, projectile and target nuclei are modelled according to the Fermi-gas ansatz. The nucleons are represented by Gaussian shaped density distributions. The wave-function of the nucleus is defined as the product wave-function of the single nucleon Gaussians. In configuration space the centroids of the Gaussians are randomly distributed within a sphere.

In the present study, we use the most recent version of UrQMD transport model, UrQMD v3.4 \cite{urqmd3.4}. In this UrQMD model we have applied a grouping algorithm on the nucleons in the initiation part and compared the results with that from the original unmodified distributions by using the method of Principal Component Analysis (PCA).

\section{Principal Component Analysis (PCA)}

PCA is a technique which is being used widely in machine learning (ML) for pre-processing of the data as it removes the correlated features in the input variables and reduces the running time \cite{pca-ml}. This technique reduces a multi-dimensional feature space into lower dimension keeping the essential features nearly intact. This is an unsupervised statistical method by which the inter-relations between the input variables are examined and newer variables called principal components (PCs) are extracted. The new variables are represented by orthogonal eigenvectors and eigenvalues giving the variables and the corresponding variances respectively. It is seen that a large fraction of the total eigenvalues are contained in the first few components thereby reducing the original dimension of the data space into only those few dimensions. It should be noted that even though the original input variables have physical interpretations, the newer variables (PCs) being a combination or derived from the input variables might not be physically interpretable. Proper combinations might be attempted to obtain physically meaningful variables from the PCs \cite{pca-physical-variable}. The main advantage is to reduce the computing complexity and extract the desired features. In this exercise, two steps are extremely important, first the construction of matrix representing the variables and then process to obtain the eigenvalues and eigenvectors. In our application, a ROOT-based implementation called TPrincipal \cite{pca-root} has been used for the decomposition using the singular value decomposition (SVD) method . In this work, matrix M is constructed from single particle pseudorapidity ($\eta$), azimuthal angle ($\phi$) and transverse momentum distribution from N number of independent events and each of these variables are further divided into m bins. Therefore,  a N $\times$ m rectangular matrix M is formed. Then the SVD is applied on the final state matrix M with dimension N$\times$m, which gives:
\begin{equation*}
M = X\Sigma~Z,
\end{equation*}
where X and Z are orthogonal matrices of N $\times$N and m$\times$m dimensions, respectively and $\Sigma$ is a rectangular matrix of N$\times$m dimensions. The matrix elements $\Sigma_{kk}$ of the $\Sigma$ matrix (where k refers to the row and column indices of the matrix) carry a physical meaning and they are arranged in a strictly descending order in magnitude. In the present case, distribution of a variable $f$ in i$^{th}$ event can be expressed as:
\begin{equation*}
f = \sum_{j=1}^{m} x_{j}^{i}\sigma_{j}z_{j} = \sum_{j=1}^{m} v_{j}^{i}z_{j}; i= 1,..,N
\end{equation*}
where z$_{j}$s are orthogonal vectors such that z$_{i}^{T}$z$_{j}$ = $\delta_{ij}$, $\sigma_{j}$s are the singular values of $\Sigma$ matrix, i is the running index for number of events (1..N) and v$_{j}$=x$_{j}\sigma_{j}$ is a coefficient corresponding to a eigen vector z$_{j}$. In PCA, first few $\sigma_{j}$s are relevant and they are generally sufficient to describe the variable.

In high energy heavy ion collisions, PCAs have been used primarily in analysing azimuthal correlation data first in \cite{pca-bhalerao}, where the potential of the components in obtaining the finer structures of the azimuthal correlations  was mentioned and was followed by an analysis of the CMS data using the PCA technique~\cite{pcaCMS}. A connection between the PCA eigenvalues with the event-averaged flow coefficients was attempted in \cite{pca-eigen-flow, pca-eigen-flow2}. The sensitivity in exploring the PCA eigenvalues to the initial partonic cluster structures at the initial phase was made in \cite{pca-shreyashi}.
In this work, we have studied the PCA sensitivity to the initial nucleonic arrangements in a colliding nucleus which is responsible for hot spot like structures in heavy ion collisions. Here, for the PCA analysis the columns of the input matrix in a row is constructed by binning the distributions of the input variables in an event. All the events analysed are then form all such rows forming the entire matrix.
As per PCA analysis, the PC1 represents the variance of the 1$^{st}$ decomposed component. The introduction of the grouping in the nucleonic structure is likely to cause hot spots in the initial state affecting the corresponding variance to be seen in the PCA eigenvalues.

\section{Implementation of hot spot at the nucleon level in UrQMD }
% to implement initial state fluctuation which create hot-spots
% like structures in the initial state we  incorporate a 
In the present study, we have implemented a rearrangement of nucleon positions to create hot spot like structures in the  projectile (Pb) and target (Pb) nuclei in UrQMD transport model as described below. A nucleon is selected as a seed at random and the seed is taken as the center of the hot spot. All nucleons whose inter-nucleonic distance with respect to the seed nucleon on the transverse plane lie within a certain radial distance (parameter \textit{R}), are brought closer to the seed nucleon by reducing the radial distances of the nucleons by a certain factor (parameter \textit{dR}). 
\iffalse
\begin{figure}
\centering
\begin{minipage}{.5\textwidth}
  \centering
%  \includegraphics[width=1.0\linewidth]{xy_default_b3fm.png}
  \includegraphics[width=66mm,height=5.1cm]{xy_default_b3fm.png}
%  \captionof{figure}{A figure}
  \label{fig:test1}
\end{minipage}%
\begin{minipage}{.5\textwidth}
  \centering
 \includegraphics[width=66mm,height=5.1cm]{xy_2fm_hotspot.png}
 %\captionof{figure}{Another figure}
  \label{fig:test2}
\end{minipage}
\caption{\label{xy_uncluster} The XY distributions of the nucleons in projectile and target Pb nucleus in the transverse plane with the impact parameter b =3 fm before hot spot creation (top) and after hot spot creation (bottom) with parameters \textit{R}=2 fm, \textit{dR}=0.5 and \textit{dz}=$\pm$ 0.5 fm }
\end{figure}
\fi
\begin{figure}
    \centering
    \includegraphics[width=66mm,height=5.1cm]{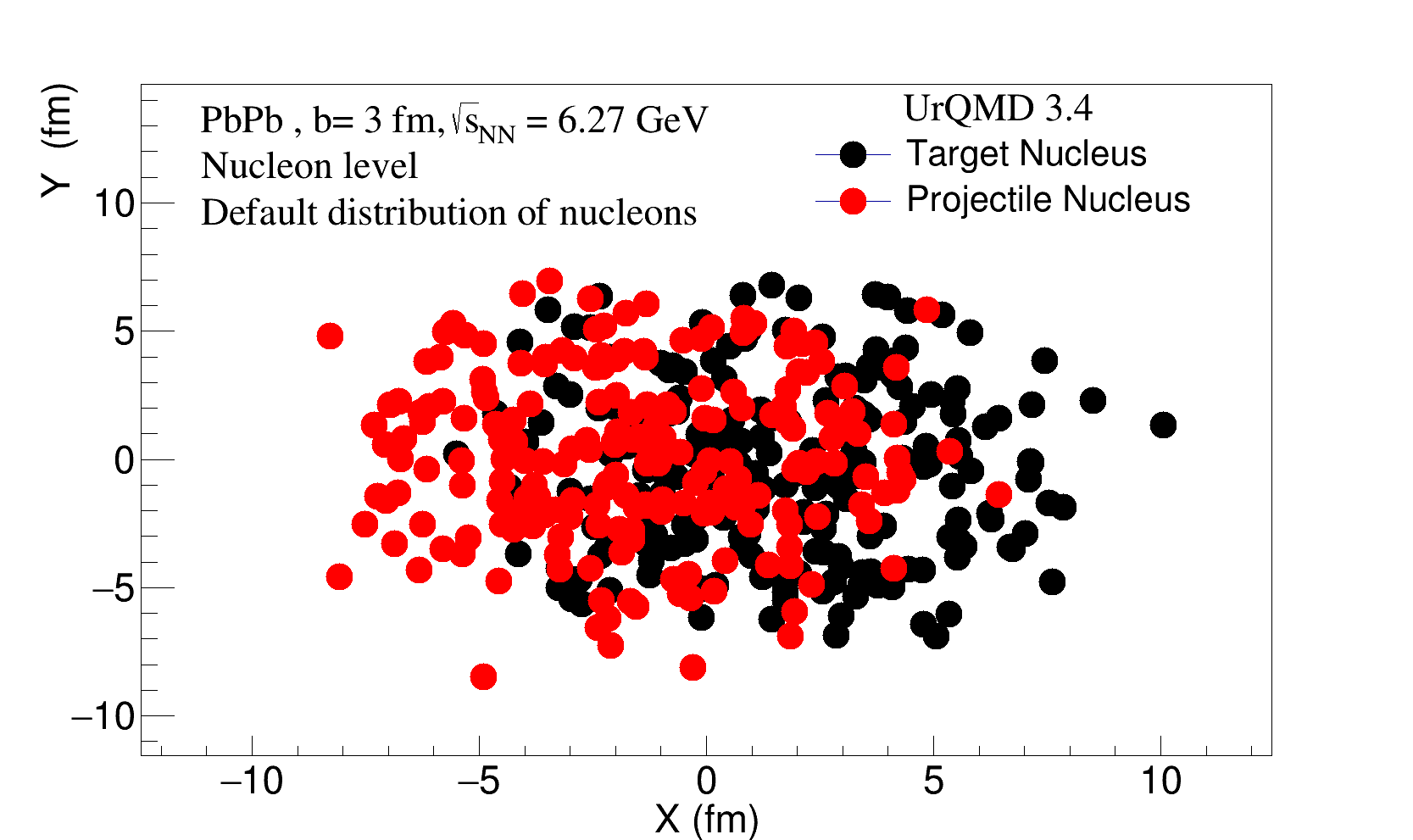}
    \includegraphics[width=66mm,height=5.1cm]{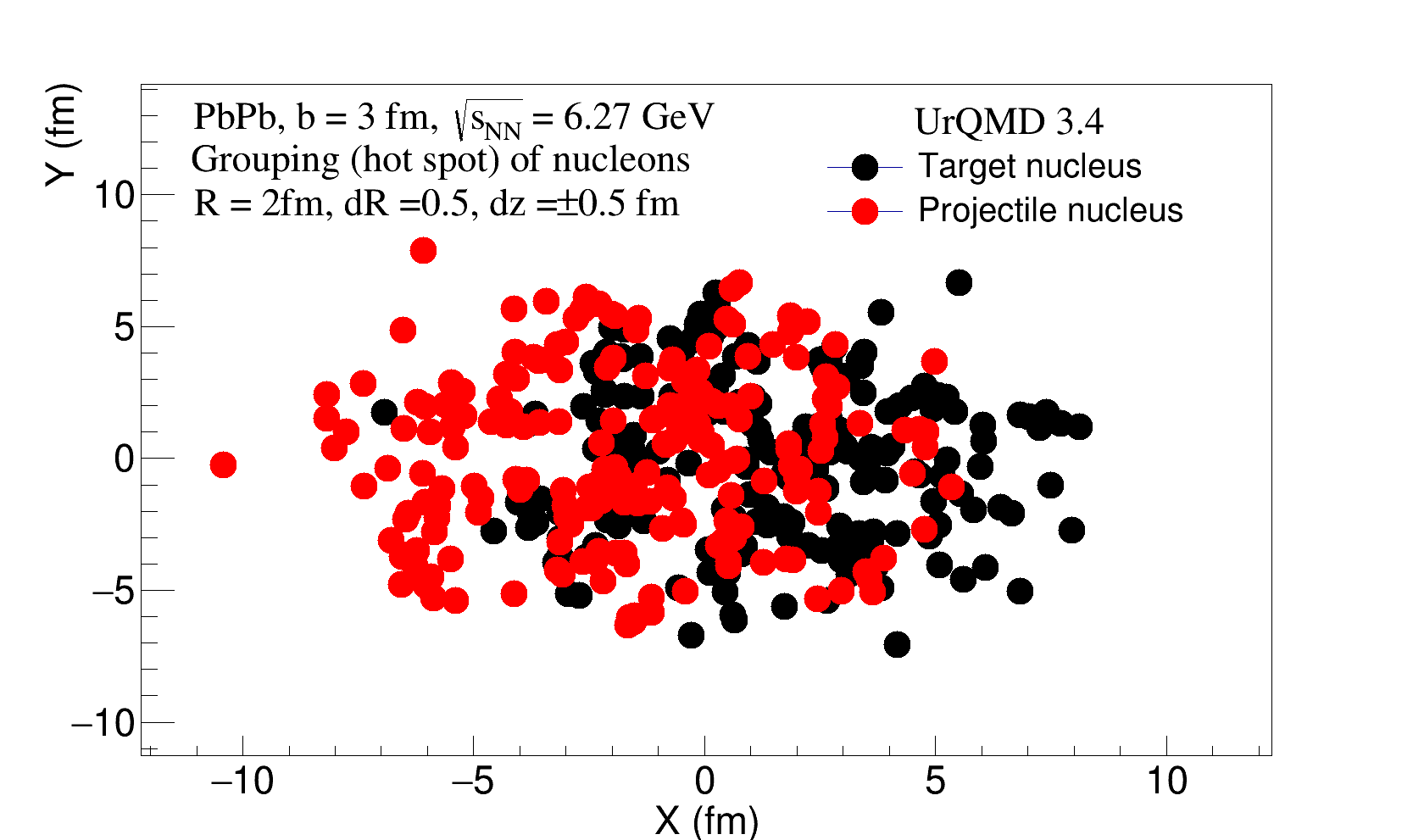}
    \caption{\label{xy_uncluster} The XY distributions of the nucleons in projectile and target Pb nucleus in the transverse plane with the impact parameter b =3 fm before hot spot creation (top) and after hot spot creation (bottom) with parameters \textit{R}=2 fm, \textit{dR}=0.5 and \textit{dz}=$\pm$ 0.5 fm}
    %\label{fig:enter-label}
\end{figure}

\begin{figure}[htbp]
	\centering\includegraphics[width=85mm,height=5.0cm]{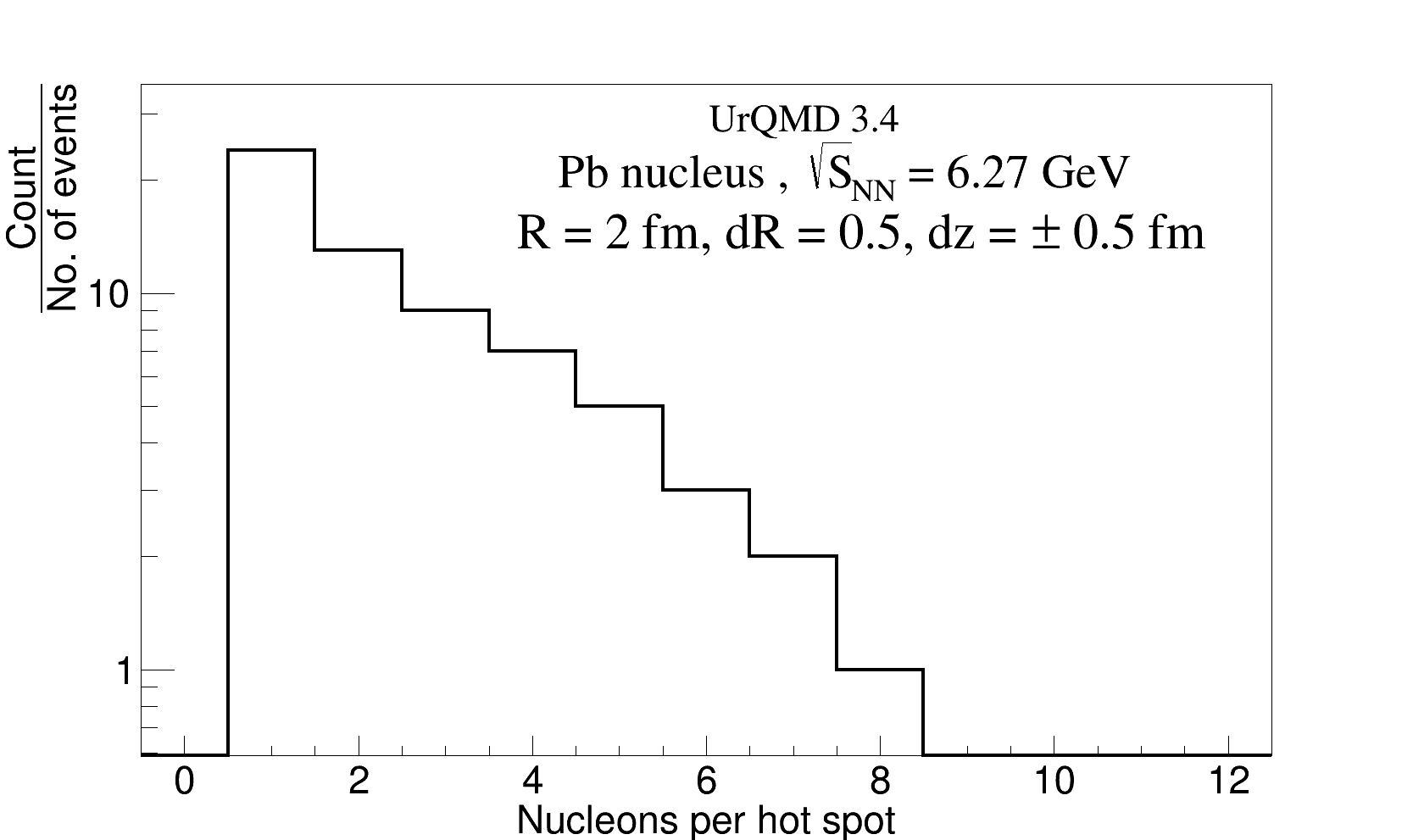}
	%	\caption{[Color online] .}
\caption{\label{nucleons_in_cluster} The number of nucleons per hot spot  with parameters \textit{R} = 2 fm , \textit{dR} = 0.5 and \textit{dZ} = $\pm$ 0.5 fm for one colliding Pb nucleus }
\end{figure}

\begin{figure}[htbp]
	\centering\includegraphics[width=85mm,height=5.0cm]{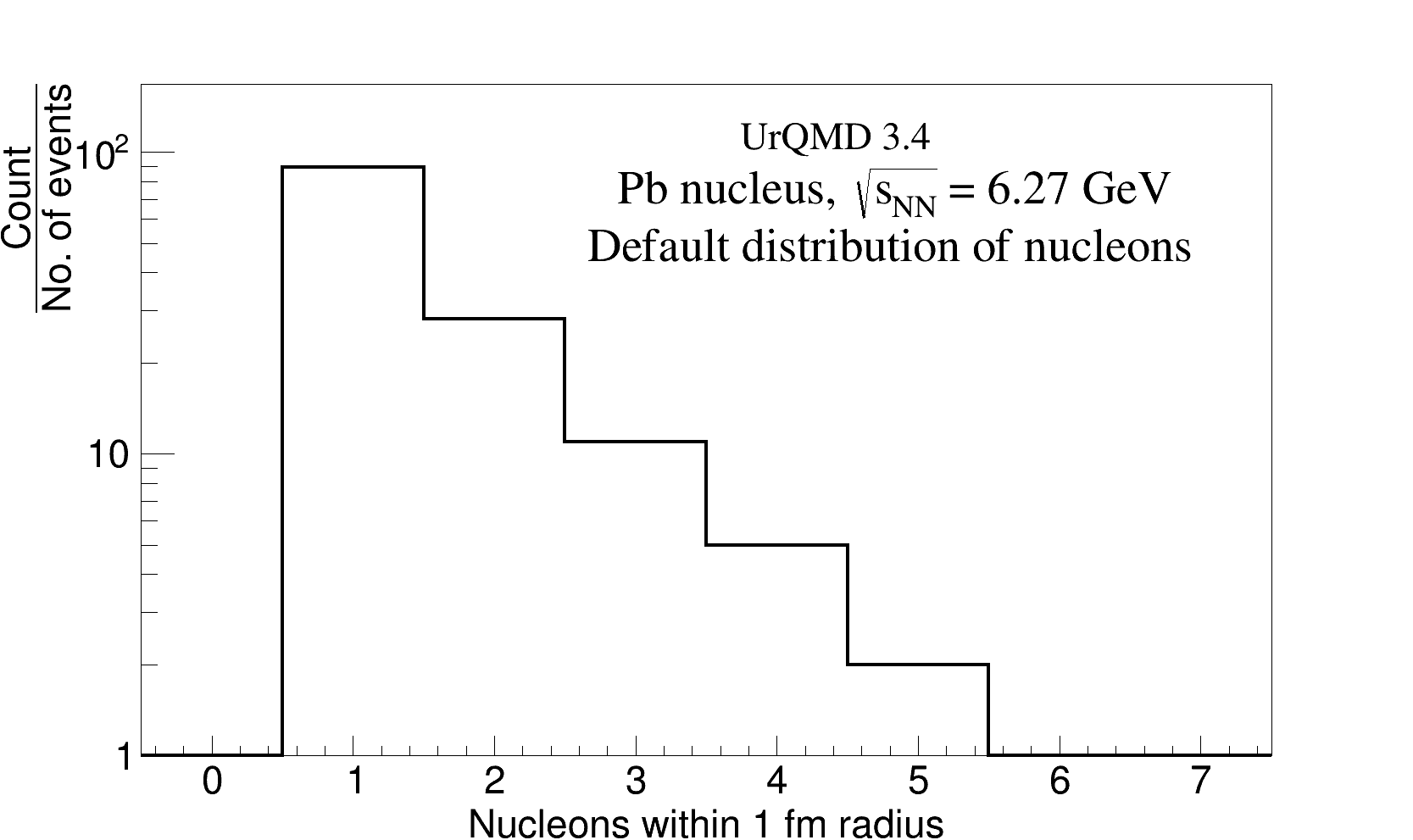}
	%	\caption{[Color online] .}
\caption{\label{nucleons_in_uncluster} The number of nucleons in R=1fm radius for one colliding Pb nucleus with default configuration }
\end{figure}

When the formation of one such localized group or say hot spot is completed, another unassigned nucleon is taken as the seed and the process continues till all nucleons are exhausted. The nucleons which are already a member of one hot spot are not considered in other hot spot. We perform the grouping of nucleons lying within $\pm$ 0.5 fm in the longitudinal (Z) direction (\textit{dZ}) with respect to the seed nucleons. Fig.~\ref{xy_uncluster} shows the XY distributions of the centre of the nucleons in an event in the projectile and the target nuclei in the transverse plane with the impact parameter b = 3 fm before hot spot creation (top) and after hot spot creation (bottom) with parameters \textit{R}=2 fm, \textit{dR}=0.5 and \textit{dz}=$\pm$ 0.5 fm. 
%We can see that before spatial rearrangement is done to create hot spots there is nearly uniform distribution of the nucleons in the colliding region and 
After the hot spot creation, we can see the modifications in the spatial arrangements of the nucleons with some local domains created in the X-Y plane. In Fig.~\ref{nucleons_in_cluster}, we have shown the average number of nucleons per hot spot with parameters \textit{R} = 2 fm , \textit{dR} = 0.5 and \textit{dZ} = $\pm$ 0.5 fm for a colliding nucleus and we can see that mostly there are single-nucleons and groups of 2 to 3 nucleons. 
In the present case, the grouping with \textit{R} = 2 fm , \textit{dR} = 0.5 corresponds to a group with maximum 1 fm radius ($\textit{R} \times \textit{dR}$). It can be seen from Fig.~\ref{nucleons_in_uncluster} that in the original distribution without clustering, mostly there are 1 or 2 nucleons within radius of 1 fm.

It is to be noted that multiplicity distribution of default distributions of nucleons in UrQMD and for different hot spot conditions with different radius parameters, R = 1, 1.5, 2, 3 fm, dR =0.5 are nearly same.
We have first studied the distributions of $\eta$, $\phi$ and $p_T$ of the produced pions as obtained from UrQMD in one (1-D) and two (2-D [$\eta-\phi$, $\eta-p_{T}$ and $\phi-p_{T}$]) dimensional spaces under various conditions of nucleonic hot spots and then compared them with the original results. 
It should be mentioned that the implementation of the hot spots in the initial configuration is expected to cause fluctuations in the final state momentum variables of the produced particles. These fluctuations are then expected to be observed also in the variance of the principal components as represented by the corresponding eigenvalues. In the present case, hot spots are introduced in the transverse plane is expected to be primarily reflected in the azimuthal distributions of the produced particles. However, other momentum components might also get affected. We have therefore investigated both 1-D and 2-D distributions.

\section{Results}\label{sec3}
In this work we have performed simulations using the UrQMD hadronic transport model with Pb+Pb collisions at $\sqrt{s}$ = 6.27 GeV. We have generated 1 million central events ( 0-10$\%$ )
to ensure that the resultant statistical errors on the event-averaged eigenvalues are not significant. We have used the produced positive and negative pions for this study. 

For the PCA analysis, we have obtained the PCA eigenvalues for both the 1-D and the 2-D distributions of the produced pions under different initial conditions characterized by the R-parameter, so that the sensitivity of PCs to the rearrangements of nucleons in the initial configurations could be tested. For the 1-D distributions, we have used $\eta$, $\phi$ and $p_T$ variables dividing each event wise distributions into 20 bins and the event by event bin wise count form the columns of the matrix for each variable. We then obtained the PCA decompositions taking all the events in the sample as rows. We have then extended the study by taking the 2-D bins of $\eta-\phi$, $\eta - p_T$ and $\phi- p_T$ divided in each case into linearized 16 bins in each event as columns of the matrix and the number of events as rows. The regions covered under this study for p$_{T}$ , $\eta$ and $\phi$ are 0 to 3 GeV/c , -1 to 1 and -$\pi$ to $\pi$ respectively.

\begin{figure}[htbp]
	 \centering
	\includegraphics[width=65mm,height=4.9cm]{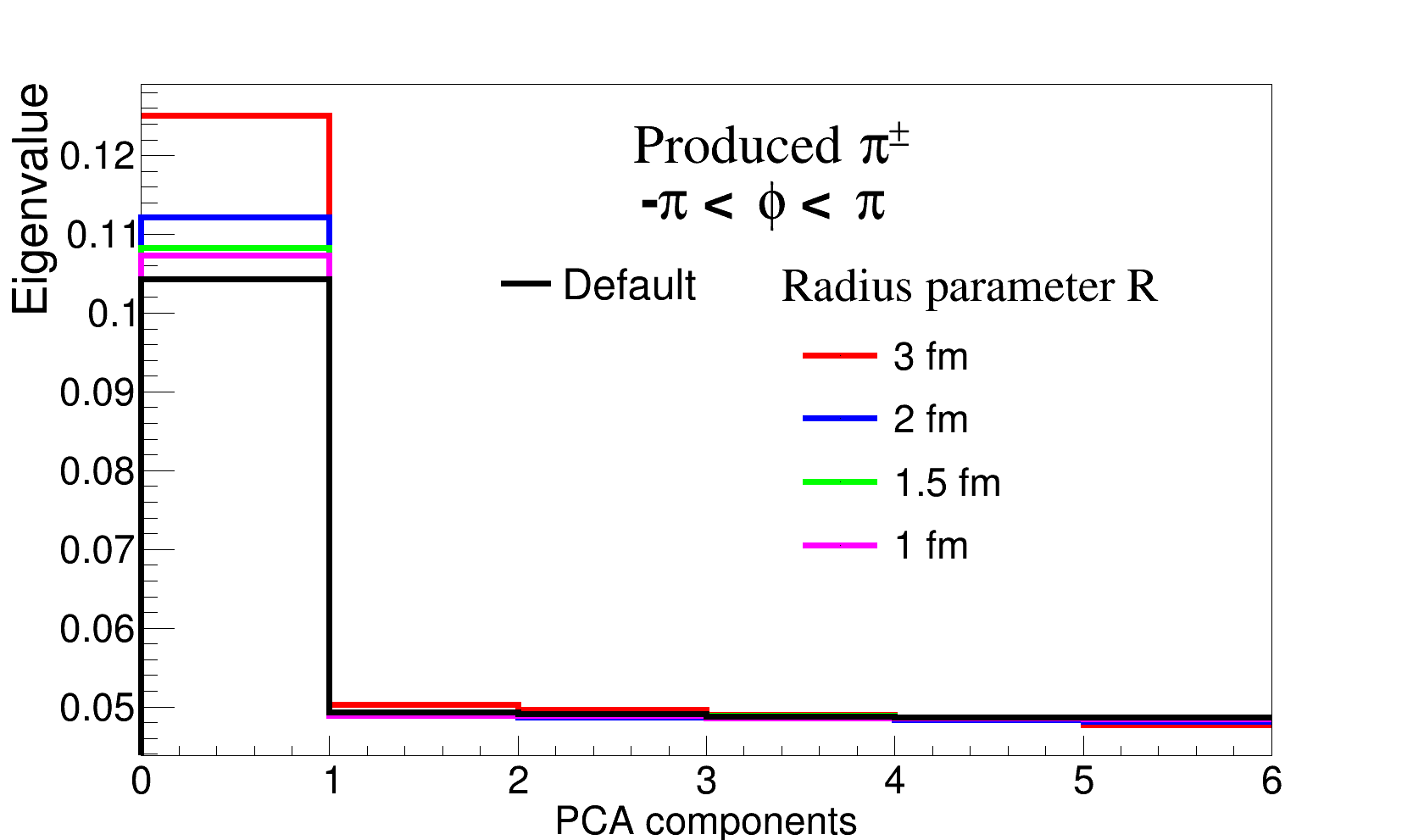}
	\includegraphics[width=65mm,height=4.9cm]{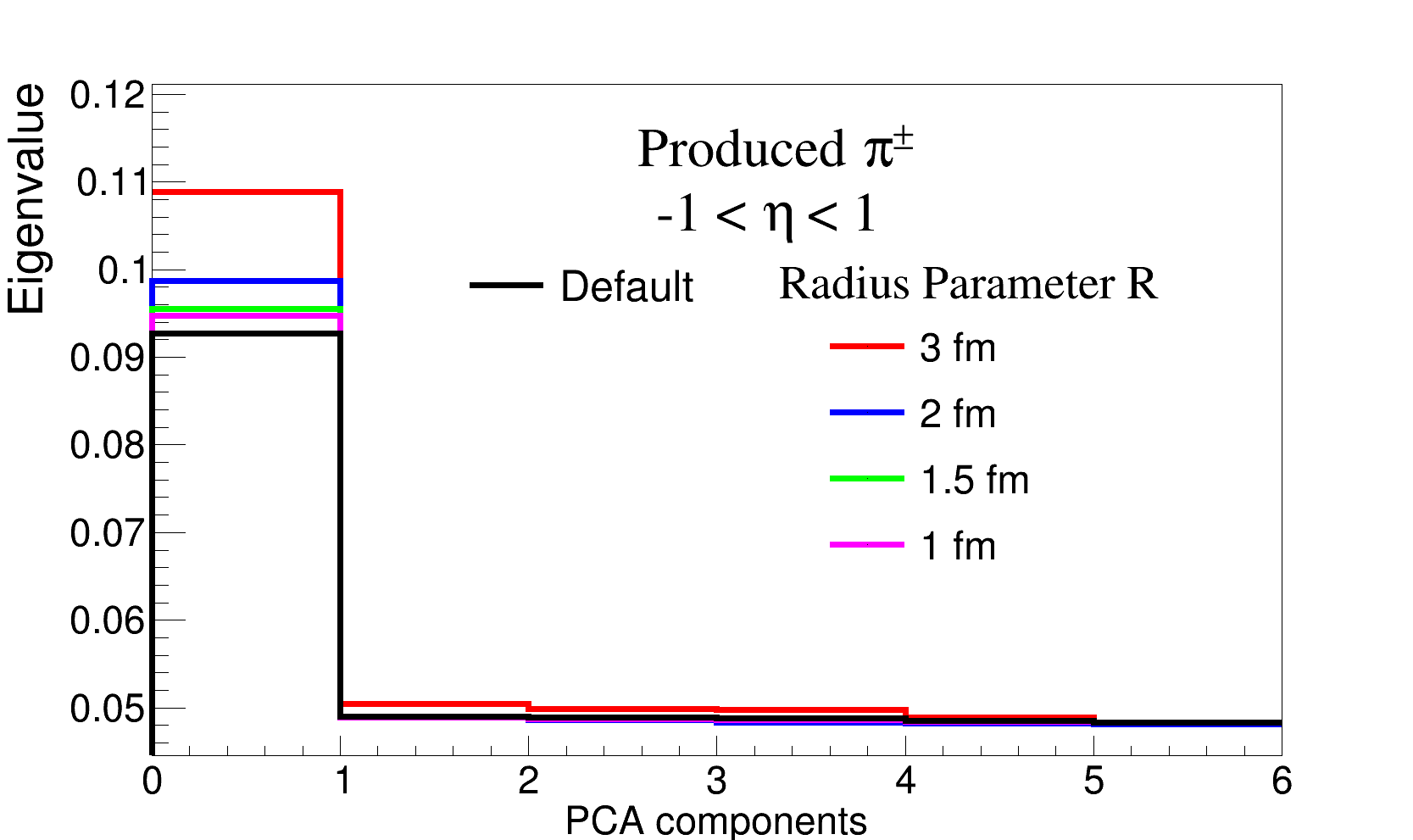}
	\includegraphics[width=65mm,height=4.9cm]{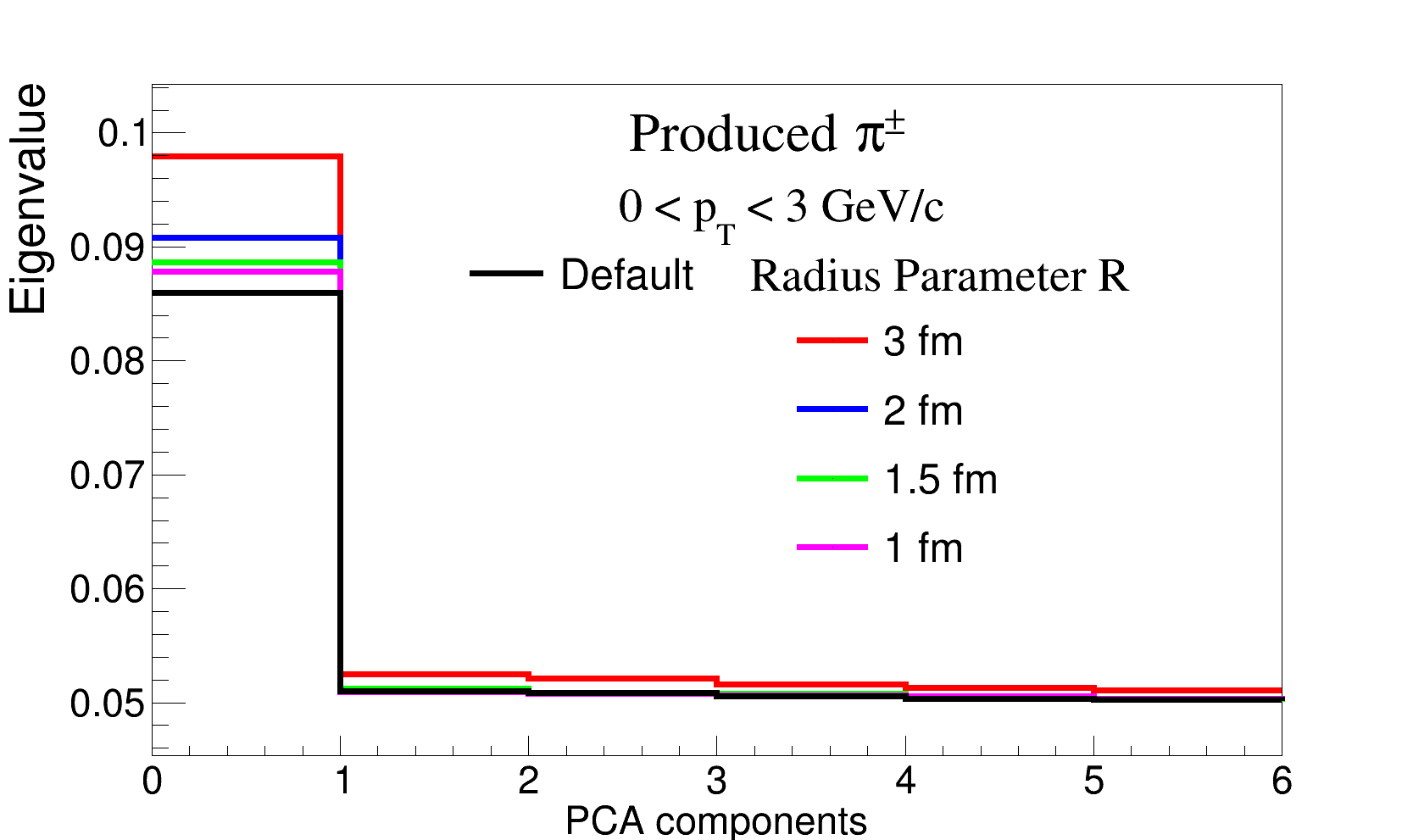}
\caption{\label{pca_etaphipt_1d} Eigenvalues in PC decomposition in 1-D distributions of $\phi$ , $\eta$ and $p_T$ bins of produced $\pi^{\pm}$ for default UrQMD and with different hot spot conditions with radius parameters, R = 1, 1.5, 2, 3 fm, dR =0.5 and dz = $\pm$ 0.5 fm  }
\end{figure}

\begin{figure}[htbp]
	 \centering
	\includegraphics[width=65mm,height=4.9cm]{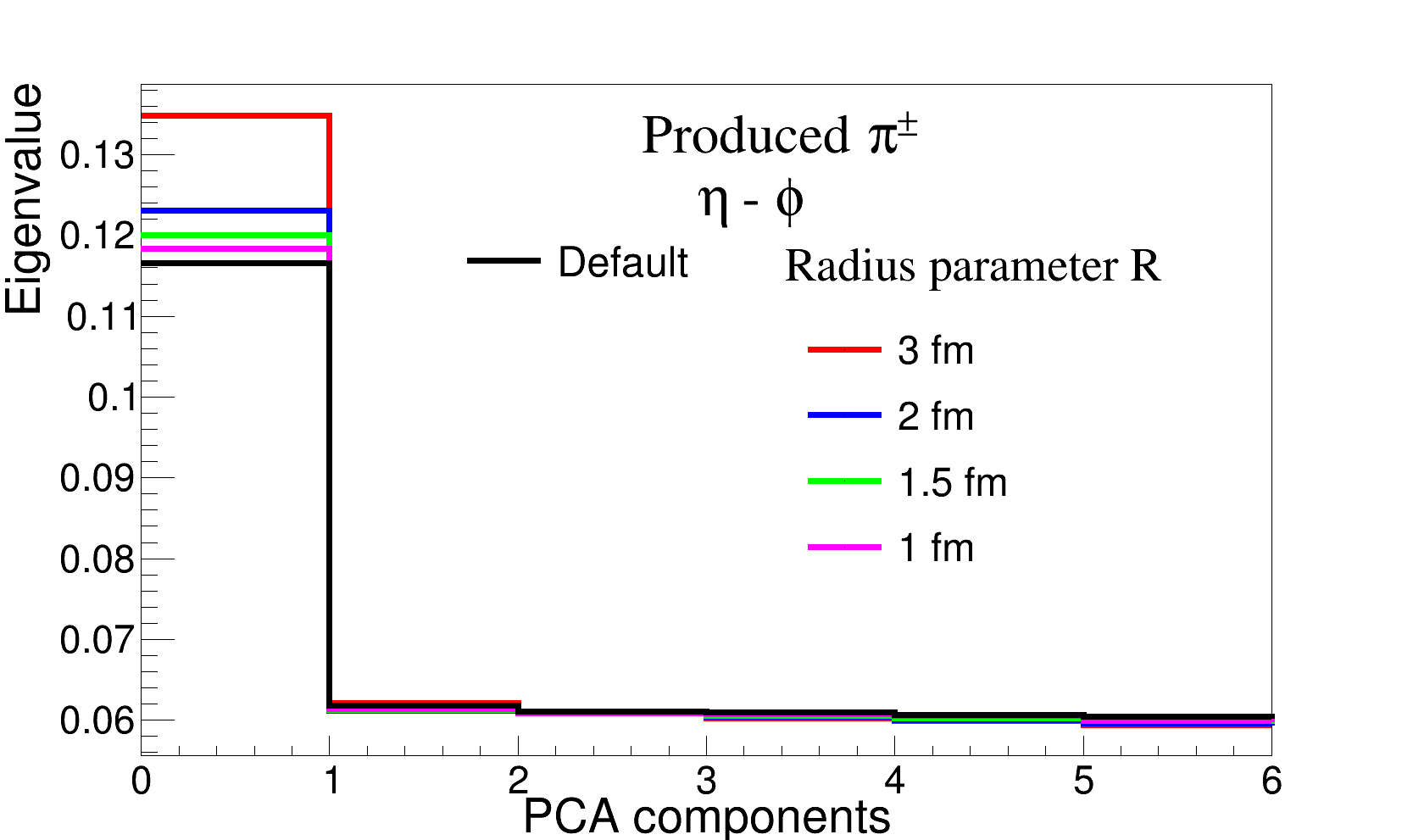}
	\includegraphics[width=65mm,height=4.9cm]{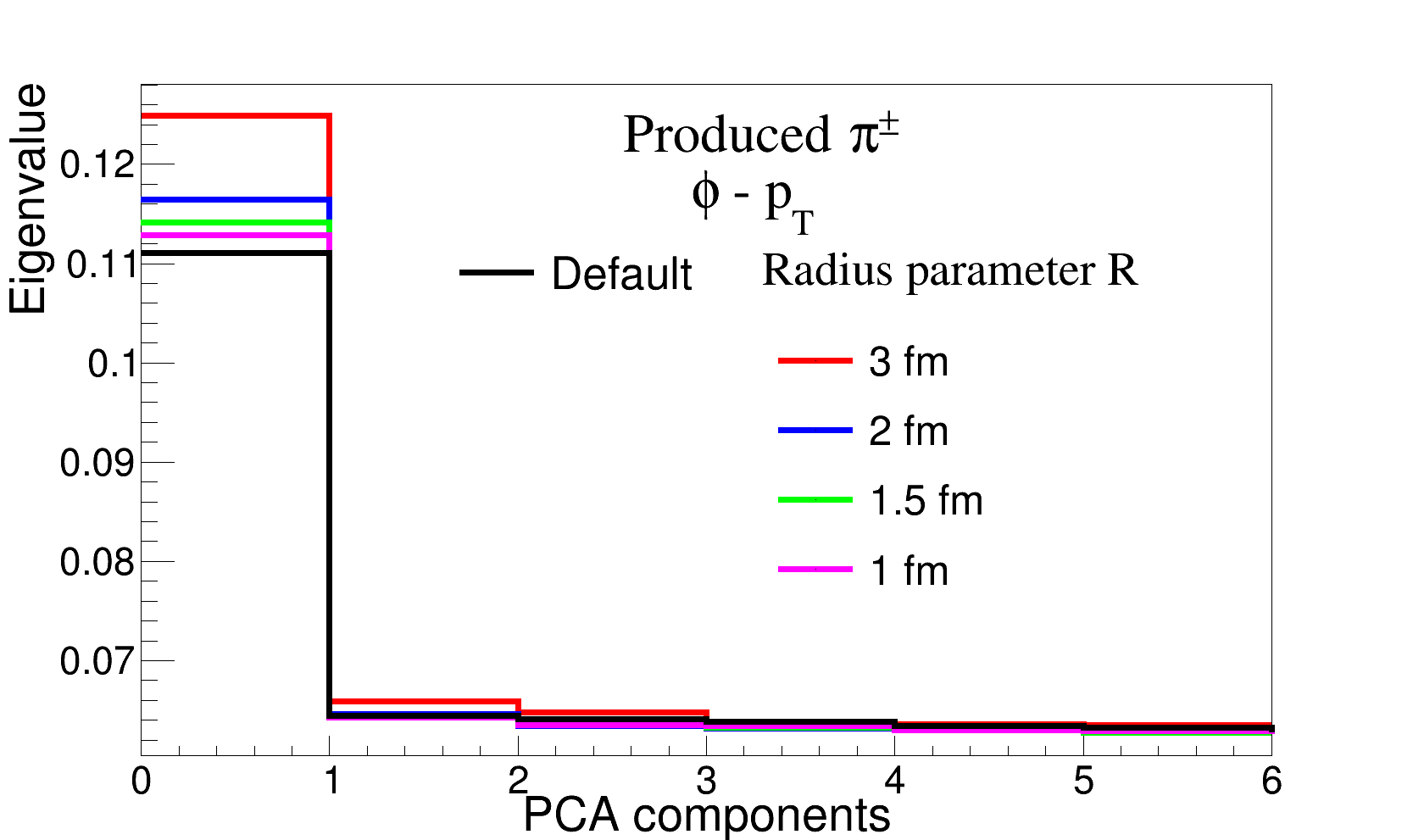}
	\includegraphics[width=65mm,height=4.9cm]{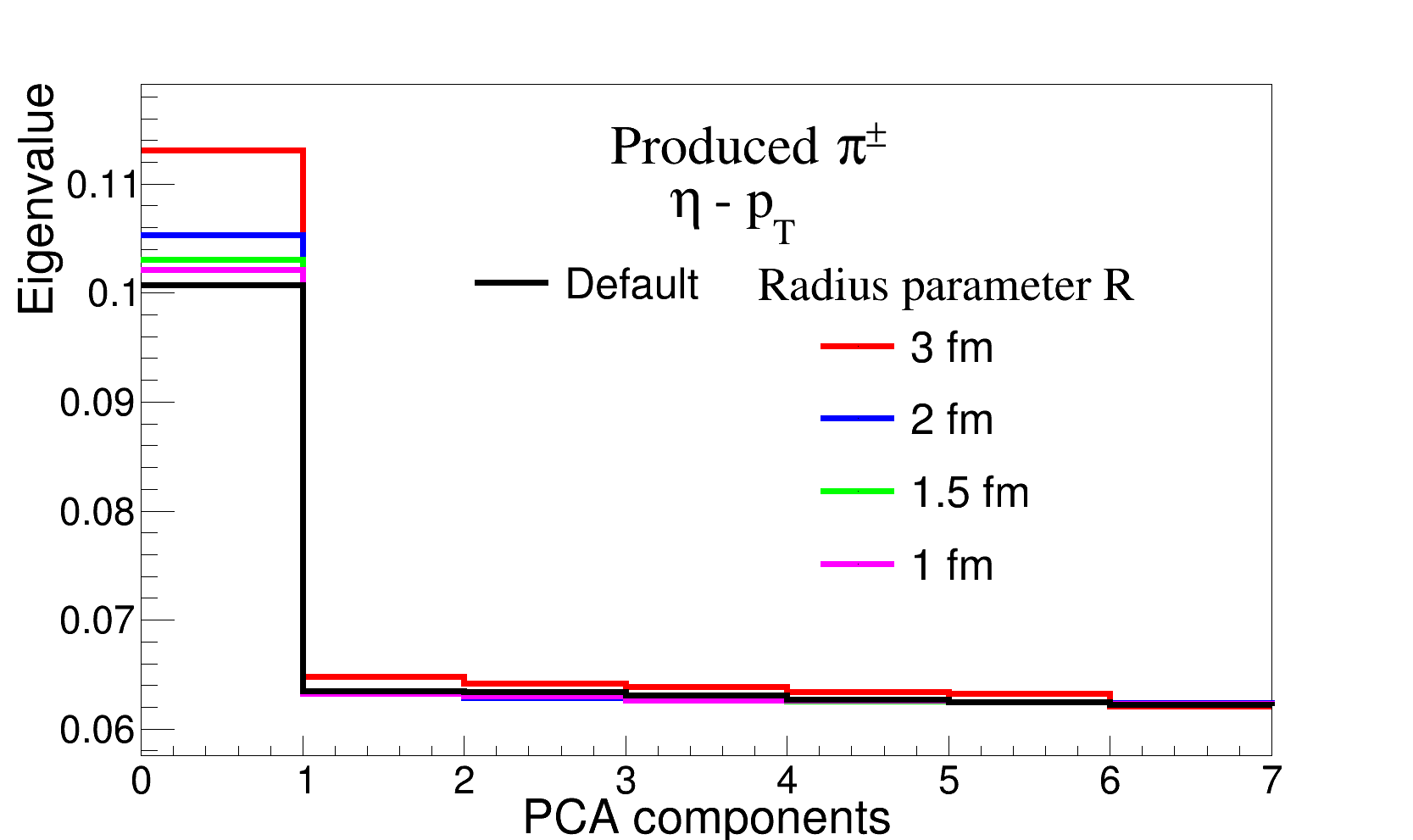}
\caption{\label{pca_etaphipt_2d} Eigenvalues in PC decomposition in 2-D distributions for ($\eta$-$\phi$), ($\phi$-$p_T$), ($\eta$-$p_T$) bins of produced $\pi^{\pm}$ for default UrQMD and with different hot spot conditions with radius parameters, R = 1, 1.5, 2, 3 fm, dR =0.5 and dz = $\pm$ 0.5 fm  }
\end{figure}

\iffalse
\begin{figure}
\centering
\begin{minipage}{.5\textwidth}
  \centering
	\includegraphics[width=65mm,height=4.9cm]{pca_phi_1d.png}
	\includegraphics[width=65mm,height=4.9cm]{pca_eta_1d.png}
	\includegraphics[width=65mm,height=4.9cm]{pca_pt_1d.png}
%  \captionof{figure}{A figure}
%\caption{\label{pca_etaphipt}Eigenvalues in 1D for different components for $\eta, p_T $ and $\phi$ bins }
%  \label{fig:test1}
\end{minipage}
\begin{minipage}{.5\textwidth}
  \centering
	\includegraphics[width=65mm,height=4.9cm]{pca_etaphi_2d.png}
	\includegraphics[width=65mm,height=4.9cm]{pca_phipt_2d.png}
	\includegraphics[width=65mm,height=4.9cm]{pca_etapt_2d.png}
 %\captionof{figure}{Another figure}
%  \label{fig:test2}
\end{minipage}
	\caption{\label{pca_etaphipt}Eigenvalues in PC decomposition in 1-D distributions of $\phi$ , $\eta$ and $p_T$ bins (left) and for 2-D distributions for ($\eta$-$\phi$), ($\phi$-$p_T$), ($\eta$-$p_T$) bins (right) of produced $\pi^{\pm}$ for default UrQMD and with different hot spot conditions with radius parameters, R = 1, 1.5, 2, 3 fm, dR =0.5 and dz = $\pm$ 0.5 fm }
\end{figure}
\fi

 Figure~\ref{pca_etaphipt_1d} and Fig.~\ref{pca_etaphipt_2d} shows the eigenvalue distributions from the decomposition of the 1-D and 2-D distributions respectively for different values of the R parameter. 
It is seen that for all three cases of the 1-D distributions, the eigenvalues show an expected decreasing trend from PC1 to the higher components, the reduction from PC1 to PC2 is about a factor of 2 and smaller in case of higher components. 
% It is seen that for all three cases of the 1-D distributions, the eigenvalues reduce drastically after that of the 1$^{st}$ component (PC1) and remains nearly flat for the higher order components. 
The reduction in the extracted eigenvalues are moderate and PC1 does not contain a very high fraction of the total eigenvalues. This is not common in conventional PCA analysis in which the first few eigenvalues contain a major fraction. The present results could be attributed to the use of raw distributions as input matrix, in contrast to the conventional pre-processed input covariance matrix. In addition, the matrix could have noise contributions which persist even to the higher orders after decomposition.  One of the biggest advantage of this method is that the raw distributions could be used in Machine Learning as input for fast event by event processing to reveal unknown physical structures.
In the present analysis, we focussed on the highest eigenvalue which is presumably connected to a physical observable and expected to be with the highest sensitivity to the structures in data.
 In case of the 1-D and 2-D distributions, the first component (PC1) is highest and changes with the parameter R, the higher components, although non-zero, does not show appreciable change with the rearrangements in the initial configurations of the nucleon positions. We have therefore mainly analysed PC1 from both 1-D and the 2-D distributions for further study. However, the response of PC2 and PC3 have been studied as well.
 It is to be noted that the exclusion of the remaining eigenvalues might result in loss of some information \cite{pca-first-eigen}.

\begin{figure}[htbp]
		\centering
	\includegraphics[width=85mm,height=5.0cm]{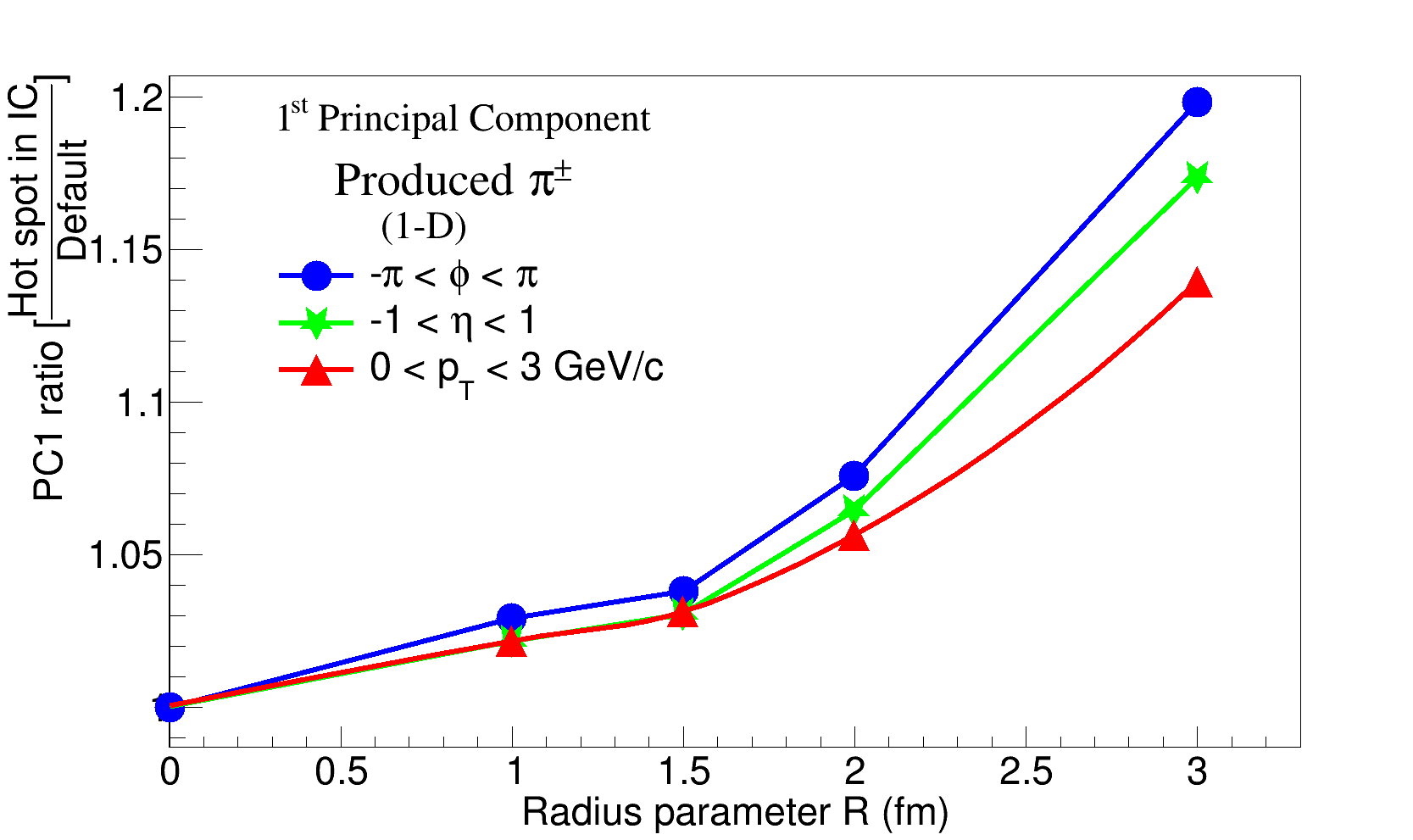}	
	%	\caption{[Color online] .}
	\caption{\label{pca-ratio-1d} First eigenvalue  (PC1) ratio of hot spot condition to the default UrQMD condition of produced $\pi^{\pm}$ with radius parameter R for  $\phi$, $\eta$ and $p_T$ (1-D) distributions.  }
\end{figure}

We have studied the PC1 values for different initial configurations generated by varying the \textit{R} parameter, i.e, R = 1 fm, 1.5 fm, 2 fm and 3 fm and compared with that for the unmodified condition for the $\phi$, $\eta$ and  p$_{T}$ distribution as shown in the Fig.~\ref{pca_etaphipt_1d} $\&$ Fig.~\ref{pca_etaphipt_2d}. From Fig.~\ref{pca_etaphipt_1d} $\&$ Fig.~\ref{pca_etaphipt_2d}, it is also clear that 1$^{\mathrm{st}}$ PC component which contains the maximum information is very much sensitive to the R-parameter or in other words to the size of the hot spot.

To evaluate the relative sensitivity we have taken ratios of the PC1 corresponding to different R-values with respect to the unmodified distribution. Fig.~\ref{pca-ratio-1d} shows the variation of the PC1 ratio with respect to the radius parameter (R) for all three 1-D cases superposed with each other. In the figures, R = 0 corresponds to unmodified original UrQMD distribution. It is seen that up to R = 2 fm, the eigenvalues increase up to 10\% with somewhat less separation between $\eta$, $\phi$, p$_{T}$ and then rises upto 20\% for R parameter of 3 fm with wider separation for three variables.
%It might be mentioned that the average number of nucleons participating in a hot spot is around 3 and 5 for R = 2 fm and 3 fm respectively.

\begin{figure}[htbp]
		\centering
	\includegraphics[width=85mm,height=5.0cm]{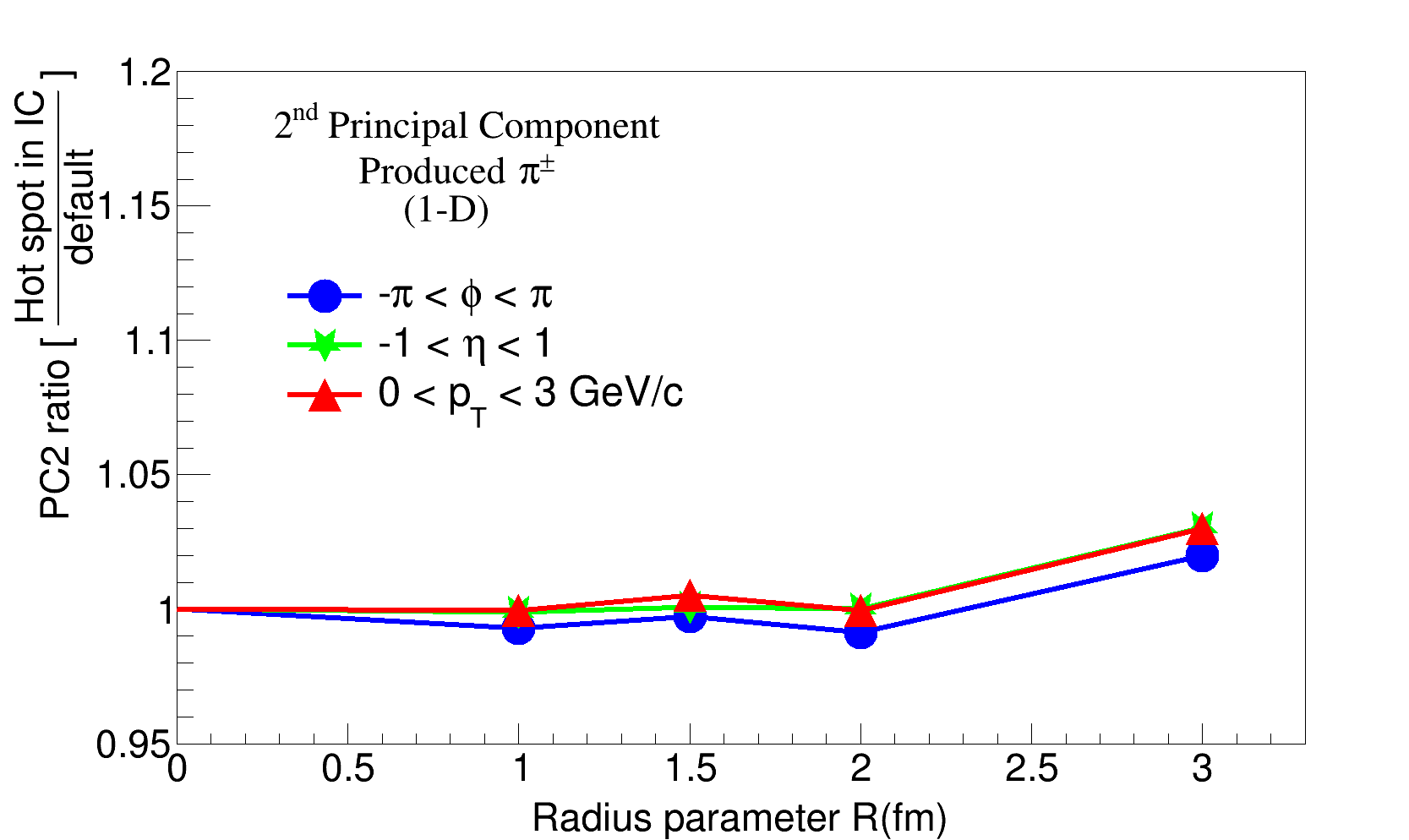}
    \includegraphics[width=85mm,height=5.0cm]{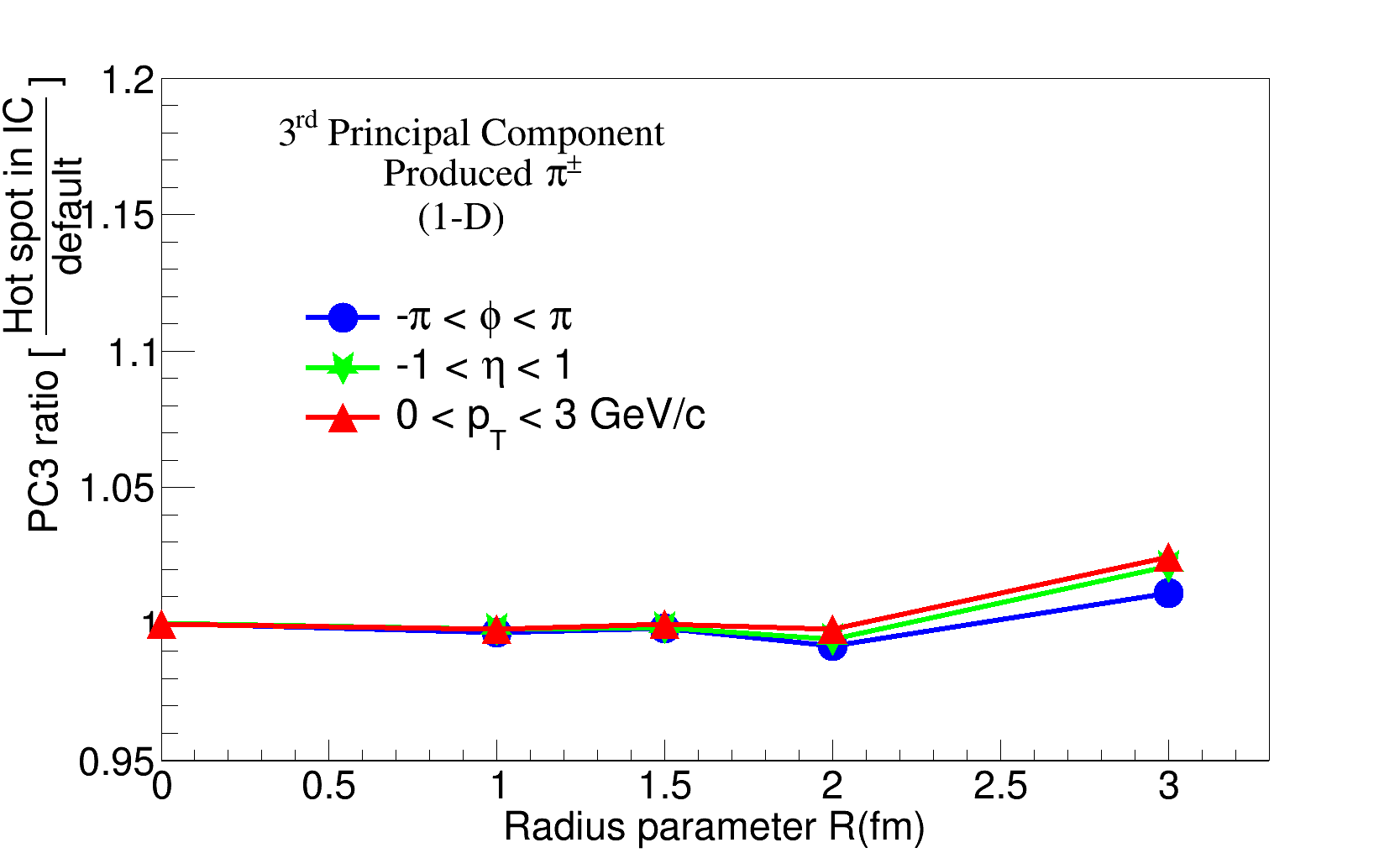}
	%	\caption{[Color online] .}
	\caption{\label{pc2-ratio-1d} (Top) Second eigenvalue (PC2) and (bottom) third eigenvalue (PC3) ratio of hot spot condition to the default UrQMD condition of produced $\pi^{\pm}$ with radius parameter R for  $\phi$, $\eta$ and $p_T$ (1-D) distributions.  }
\end{figure}

It is also observed that the azimuthal distribution shows higher sensitivity in terms of the relative increase in PC1, which is more prominent at the higher R. This could be due to the reason that the implementation of the grouping of nucleons is primarily on the azimuthal plane. It is however clear that the azimuthal nucleonic grouping affect the other momentum components i.e., $\eta$ and $p_T$ of the final state particles as well.

We have also studied the relative sensitivity of second and third principal component i.e, PC2 and PC3 for different R-values relative to unmodified case as shown in Fig.~\ref{pc2-ratio-1d}. It is to be noted that PC2 and PC3 exhibit much reduced sensitivity than PC1. While the relative change in PC1 as a function of radius parameter, R goes upto 20\% at R=3 for $\phi$, the change in PC2 and PC3 is even less than 2\% at R=3. 

\begin{figure}[htbp]
	\centering
	\includegraphics[width=85mm,height=5.0cm]{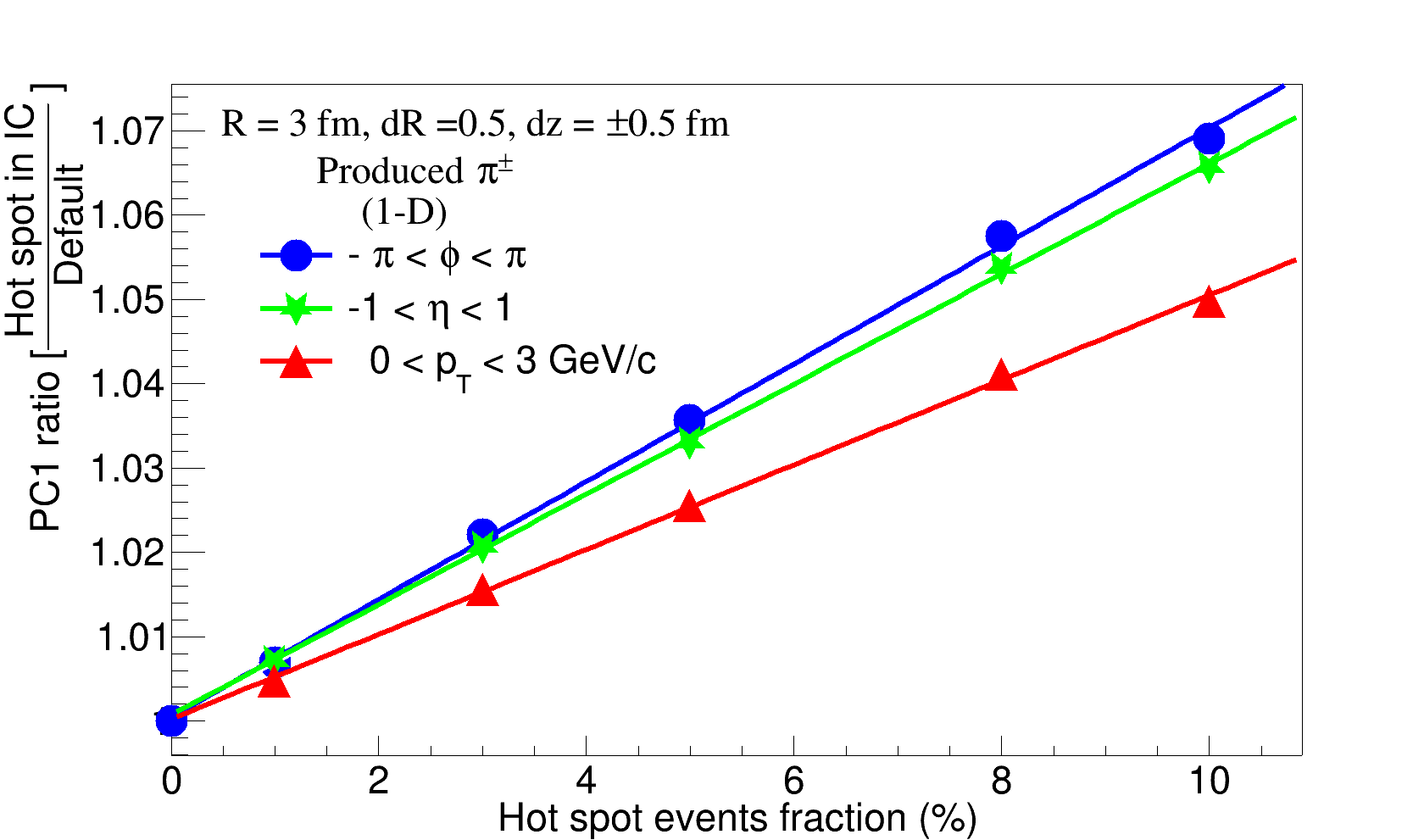}
%	\caption{[Color online] .}
	\caption{\label{pca-percentage-1d} Variation of the first eigenvalue (PC1) ratio of hot spot to the default condition with the event fraction having hot spot like events in  $\phi$, $\eta$ and $p_T$ 1-D distributions  }
\end{figure}

\begin{figure}[htbp]
	\centering
	\includegraphics[width=85mm,height=5.0cm]{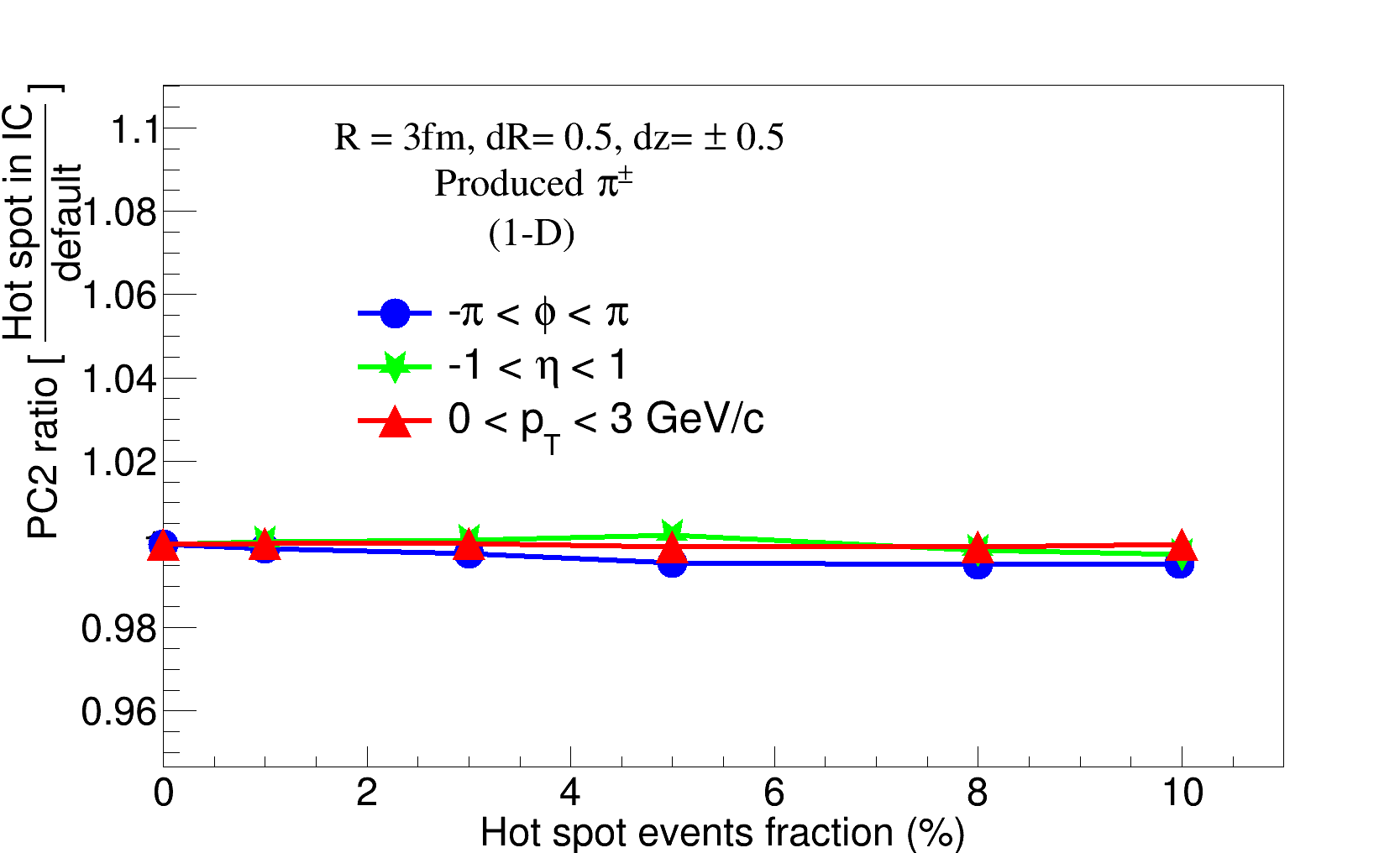}
%	\caption{[Color online] .}
	\caption{\label{pca2-percentage-1d} Variation of the second eigenvalue (PC2) ratio of hot spot to the default condition with the event fraction having hot spot like events in  $\phi$, $\eta$ and $p_T$ 1-D distributions  }
\end{figure}

\begin{figure}[htbp]
	\centering
	\includegraphics[width=85mm,height=5.0cm]{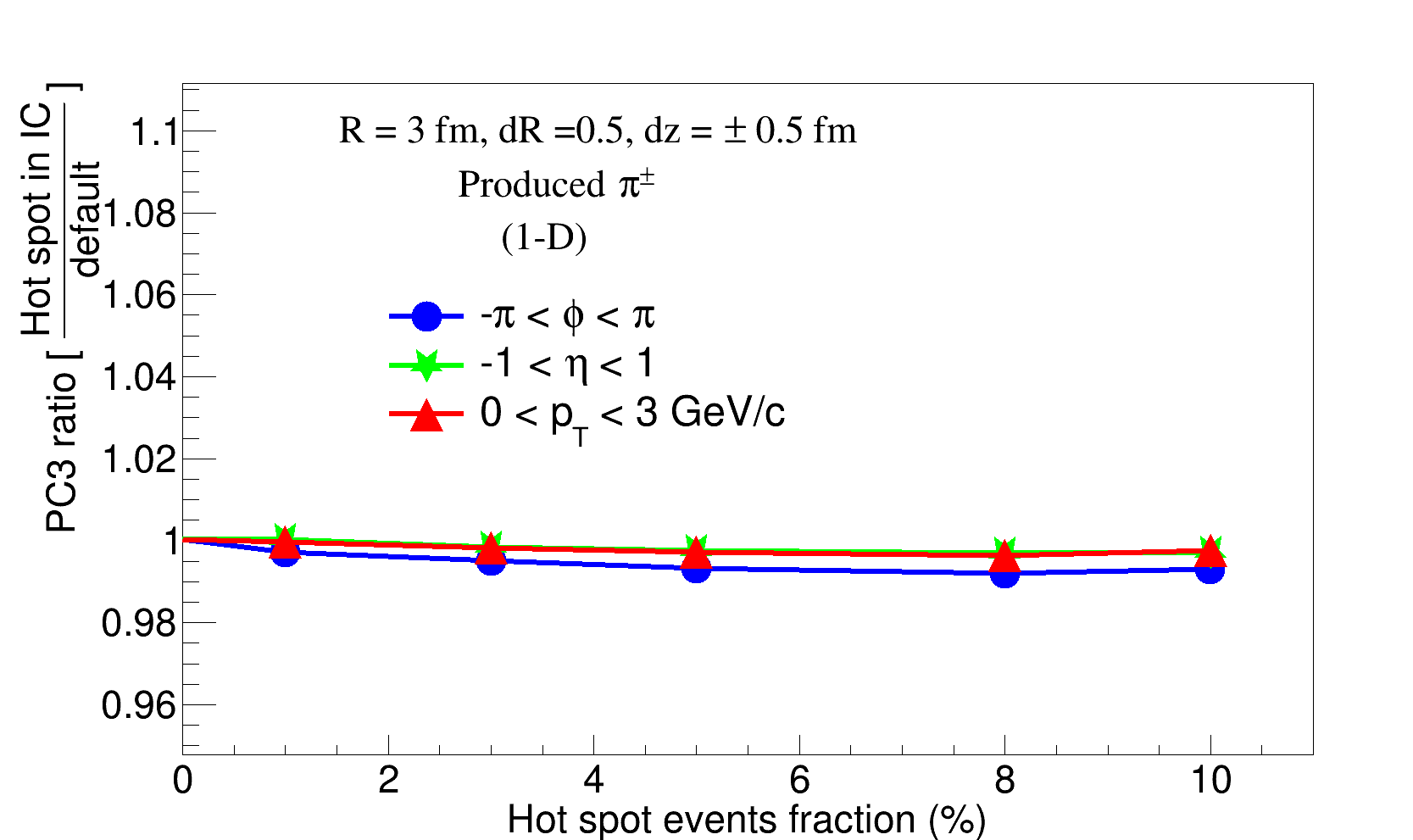}
%	\caption{[Color online] .}
	\caption{\label{pca3-percentage-1d} Variation of the third eigenvalue (PC3) ratio of hot spot to the default condition with the event fraction having hot spot like events in  $\phi$, $\eta$ and $p_T$ 1-D distributions  }
\end{figure}

In the study so far, modifications to the initial nucleon position distribution with specific parameters were introduced in every event. However, it is likely that the events with such hot-spot-like nucleon distribution will not occur in every event. In view of that, we have introduced nucleonic hot spots randomly in a varying fraction of events. Fig.~\ref{pca-percentage-1d} shows the variation of PC1 ratio with respect to the percentage of event with hot-spot with radius parameter R = 3 fm , dR =0.5 and dz = $\pm$ 0.5 fm. It is seen that for an increase in an event fraction with hot-spot upto 10\%, the event averaged PC1 increases linearly for all three distributions. The slope shows somewhat slower increase for the $\eta$ and $p_{T}$ distributions as compared to that for the azimuthal distributions. 

We have also studied the same for PC2 and PC3. As can be seen from Fig.~\ref{pca2-percentage-1d} and Fig.~\ref{pca3-percentage-1d}, response of PC2 and PC3 to the fraction of events with hot spots in the initial condition is nearly flat and not sensitive to the change.

\begin{figure}[htbp]
	\centering
	\includegraphics[width=85mm,height=5.0cm]{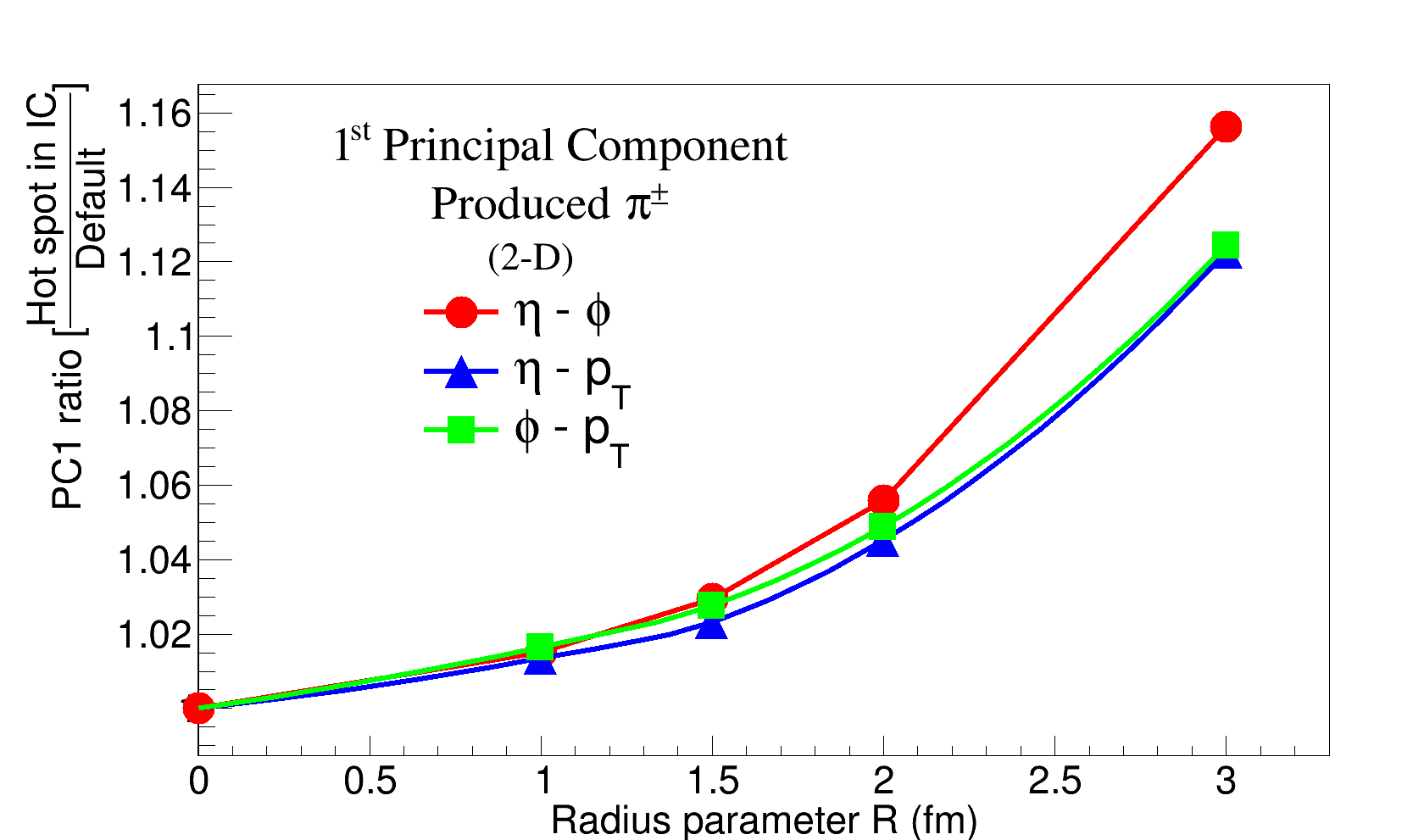}
%	\caption{[Color online] .}
	\caption{\label{pca-ratio-2d}  First eigenvalue  (PC1) ratio of hot spot condition to the default UrQMD condition of produced $\pi^{\pm}$ with radius parameter R for the decomposition of the 2-D distributions of   $\eta$-$\phi$,  $\eta$-$p_T$ and $\phi$-$p_T$.}
\end{figure}

\begin{figure}[htbp]
	\centering
	\includegraphics[width=85mm,height=5.0cm]{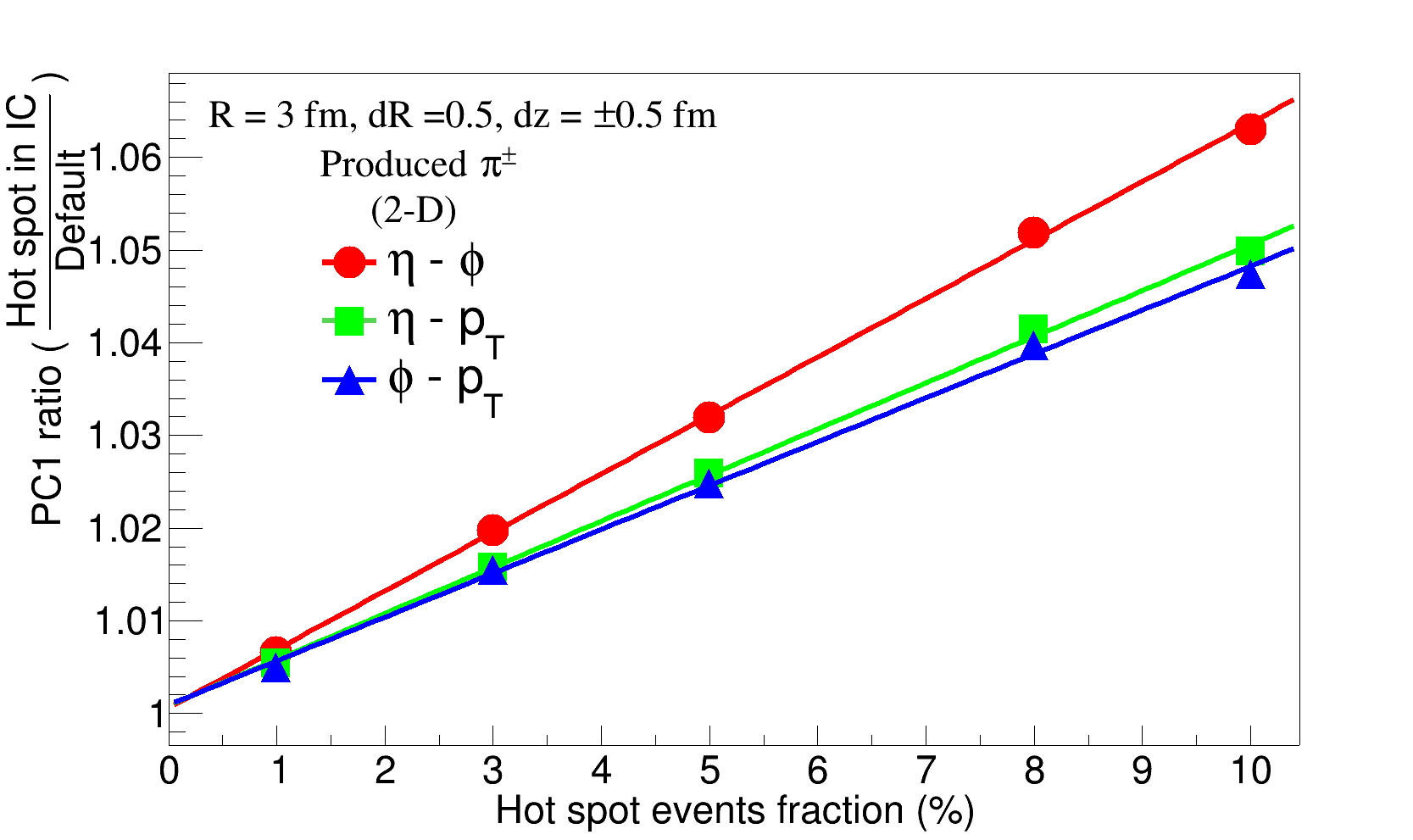}
%	\caption{[Color online] .}
	\caption{\label{pca-percentage-2d}  Variation of the first eigenvalue (PC1) ratio of hot spot to the default condition with the event fraction having hot spot like events in decomposition of the 2-D distributions in $\eta$-$\phi$,  $\eta$-$p_T$ and $\phi$-$p_T$.}
\end{figure}

The relative variation of the PC1 with respect to that unmodified configuration for the 2-D distributions is similar to those observed in 1-D cases as shown by the variation of the PC1 ratio with respect to R and the fraction of events with hot spot in initial configuration as shown in Fig.~\ref{pca-ratio-2d} and Fig.~\ref{pca-percentage-2d} respectively. In the case of 2-D distributions, as expected from the 1-D results, the sensitivity is higher for the $(\eta-\phi)$ space. This variation is similar as that observed in case of 1-D distributions suggesting that the sensitivity does not improve much due to finer structures. 
We have also checked the response of PC2 and PC3 from 2-D distributions to the variation of radius parameter, R and event fraction. Similar to 1-D, PC2 and PC3 does not exhibit any significant variation from unity with the change in radius parameter R or event fraction containing hot spot.

 \begin{figure}[htbp]
	\centering
	\includegraphics[width=85mm,height=5.0cm]{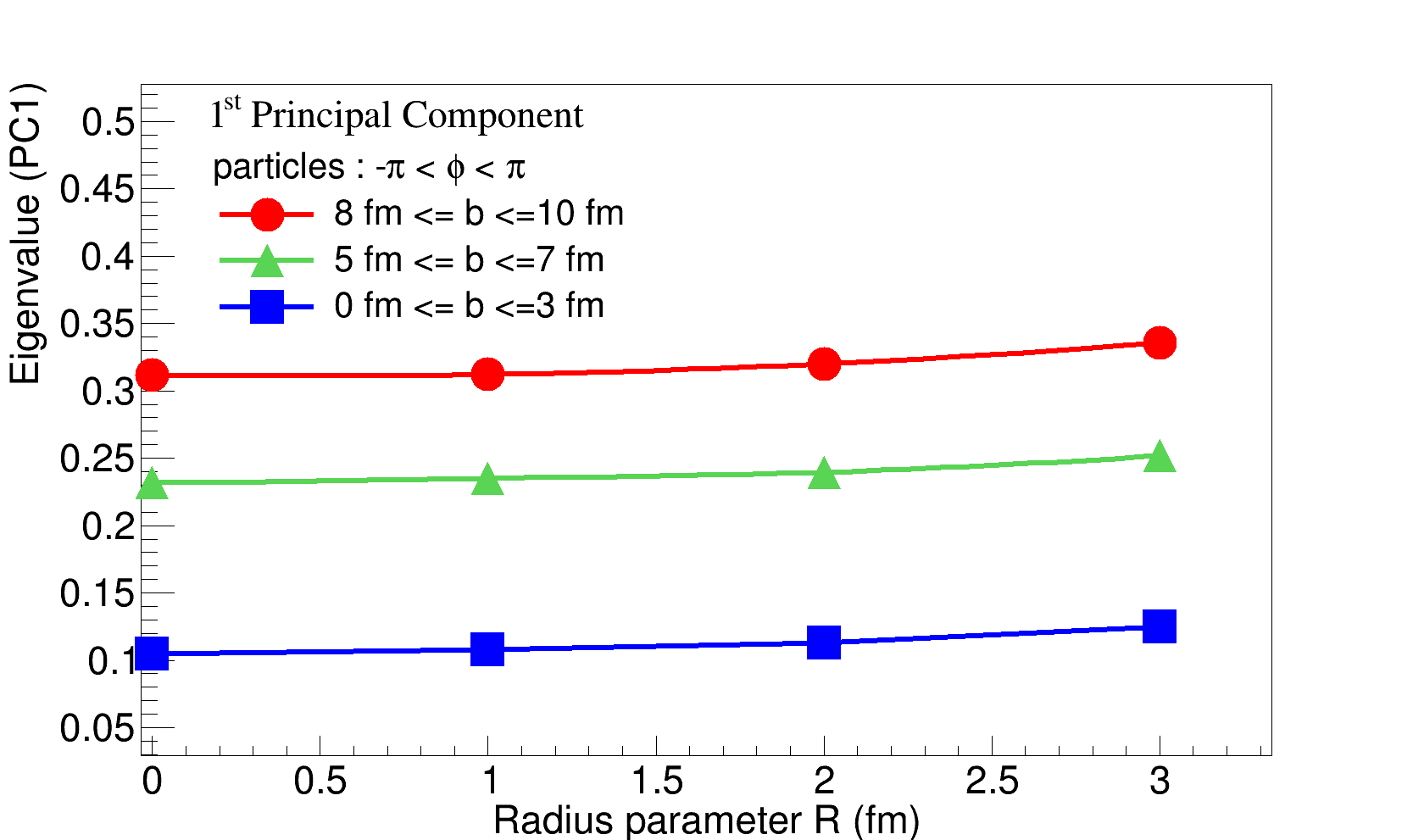}
%	\caption{[Color online] .}
	\caption{\label{pca-centrality} Variation of the first eigenvalue (PC1) with hot spot radius parameter R (R = 0 corresponds to default condition of UrQMD) for three different centralities b= 0 - 3 fm, 5-7 fm and 8-10 fm for Pb+Pb collisions in $\phi$ distribution. }
\end{figure}

 From the results so far, the sensitivity of the PC1 is highest for the $\phi$ distribution which is likely to be connected to the azimuthal asymmetry of the produced particles. From the results of the asymmetric flow studies, it is observed that the asymmetry is higher for the mid-central collisions. Keeping this in mind, it is expected that the present observables should have similar response to the centrality of the collisions.  We have therefore studied the eigenvalues for three different centralities i.e., b=0-3 fm, 5-7 fm and 8-10 fm for Pb+Pb collisions. 
 %In this study modifications to the initial nucleon positions have been introduced in all events. 
 It is found that the eigenvalues for all the 1-D distributions increase with reduced centralities. As a specific example, we have shown in Fig.~\ref{pca-centrality} the PC1 results for the azimuthal distributions at 3 different centralities in which the values increase by a factor of three from 0.1 at the most central to 0.3 at the mid-central events. The increase of PC1 for the $\phi$-distribution towards lower centrality is a reflection of the higher azimuthal asymmetry coefficients towards more non-central collisions.

\begin{figure}[htbp]
	\centering
	\includegraphics[width=85mm,height=5.0cm]{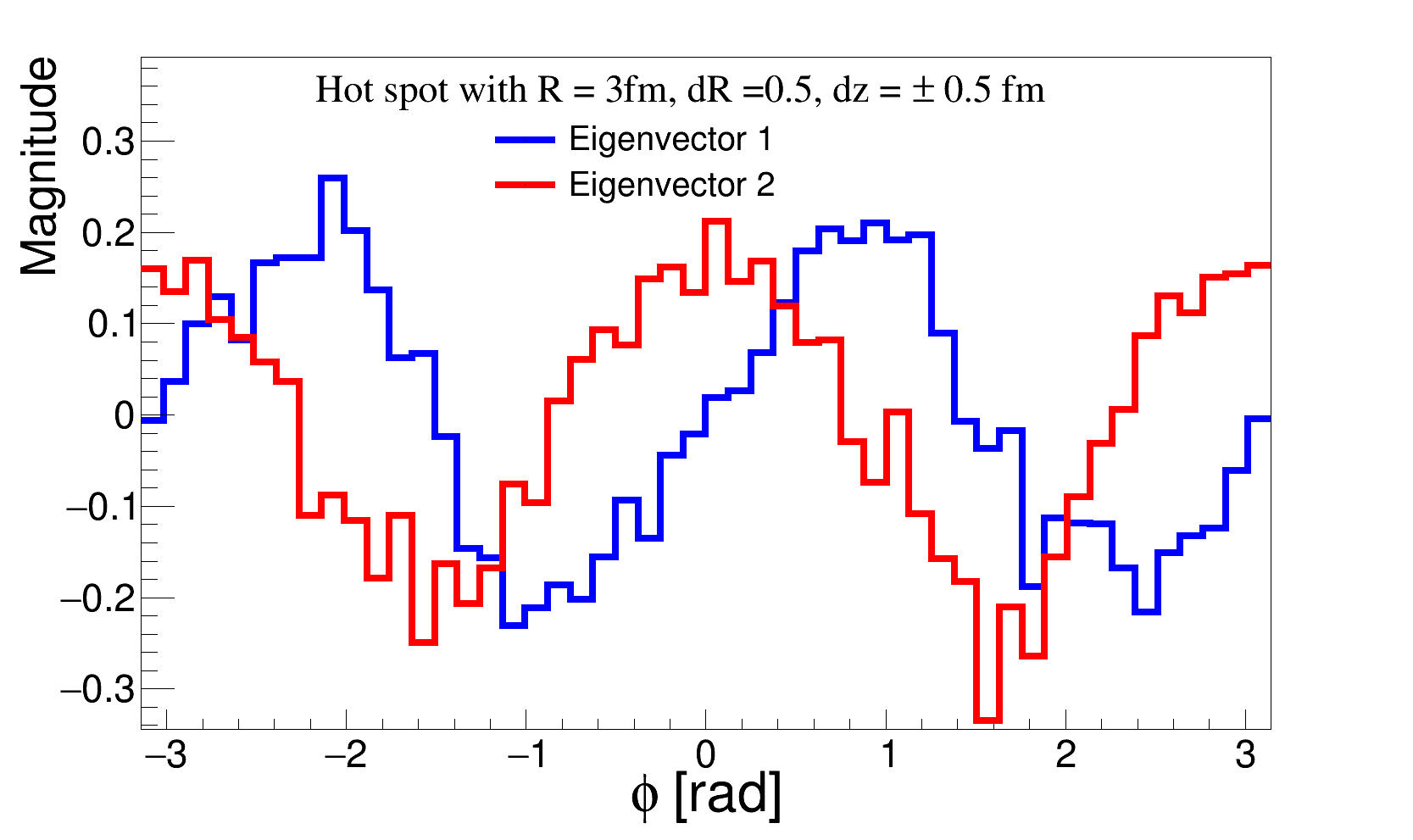}
%	\caption{[Color online] .}
	\caption{\label{eigenvector_phi}The 1$^{st}$ and 2$^{nd}$ eigenvectors obtained from PCA decomposition of $\phi$-distributions for events where hot spots are implemented }
\end{figure}

\begin{figure}[htbp]
	\centering
	\includegraphics[width=85mm,height=5.0cm]{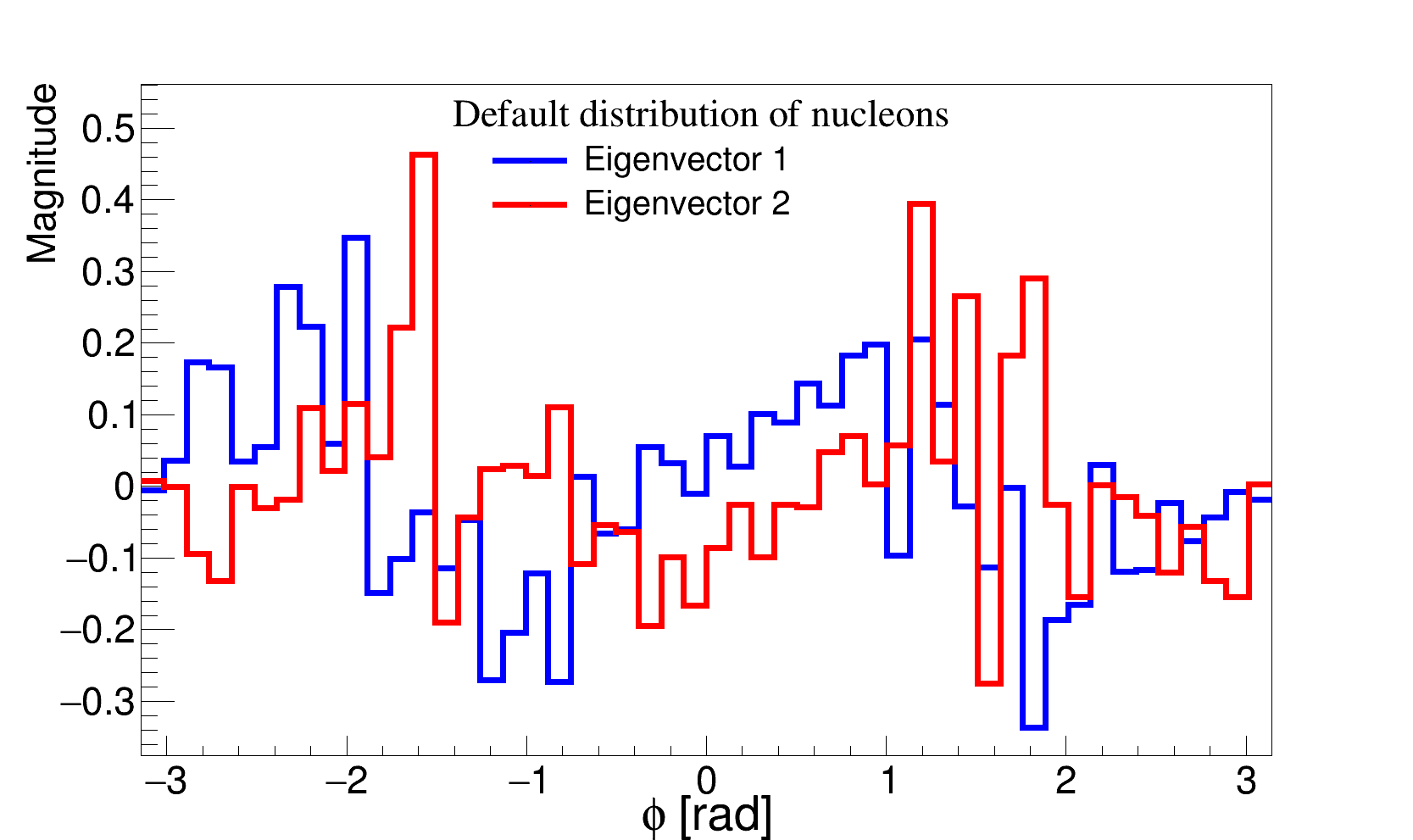}
%	\caption{[Color online] .}
	\caption{\label{eigenvector_phi_uncluster}The 1$^{st}$ and 2$^{nd}$ eigenvectors obtained from PCA decomposition of $\phi$-distributions for events where hot spots are not implemented }
\end{figure}

Additionally, We have also extracted the eigenvectors of the variables from the PCA decompositions.
Fig.~\ref{eigenvector_phi} shows the 1$^{st}$ and 2$^{nd}$ eigenvectors obtained after the PCA decomposition of $\phi$-distributions for events where hot spots are implemented
with radius parameter R = 3 fm. In heavy ion collisions, azimuthal distributions are represented in terms of Fourier harmonics~\cite{flow4} as, 
\begin{equation}
\frac{dN}{d\phi} = \frac{1}{2\pi}( 1 + 2\sum_{n}v_{n}\cos(n\phi) + 2\sum_{n}a_{n}\sin(n\phi) ).
\end{equation}

One can immediately identify the 1$^{st}$ and 2$^{nd}$ eigenvectors from PCA with Fourier decomposition bases of sin(2$\phi$) and cos(2$\phi$), respectively. We have also checked other higher components but they do not show any definite pattern which is likely to be due to smaller contributions of the higher order flow components. It is however, clear that PCA can extract the appropriate basis set that allows a more meaningful representation of a distribution.

\begin{figure}[htbp]
	\centering
	\includegraphics[width=85mm,height=5.0cm]{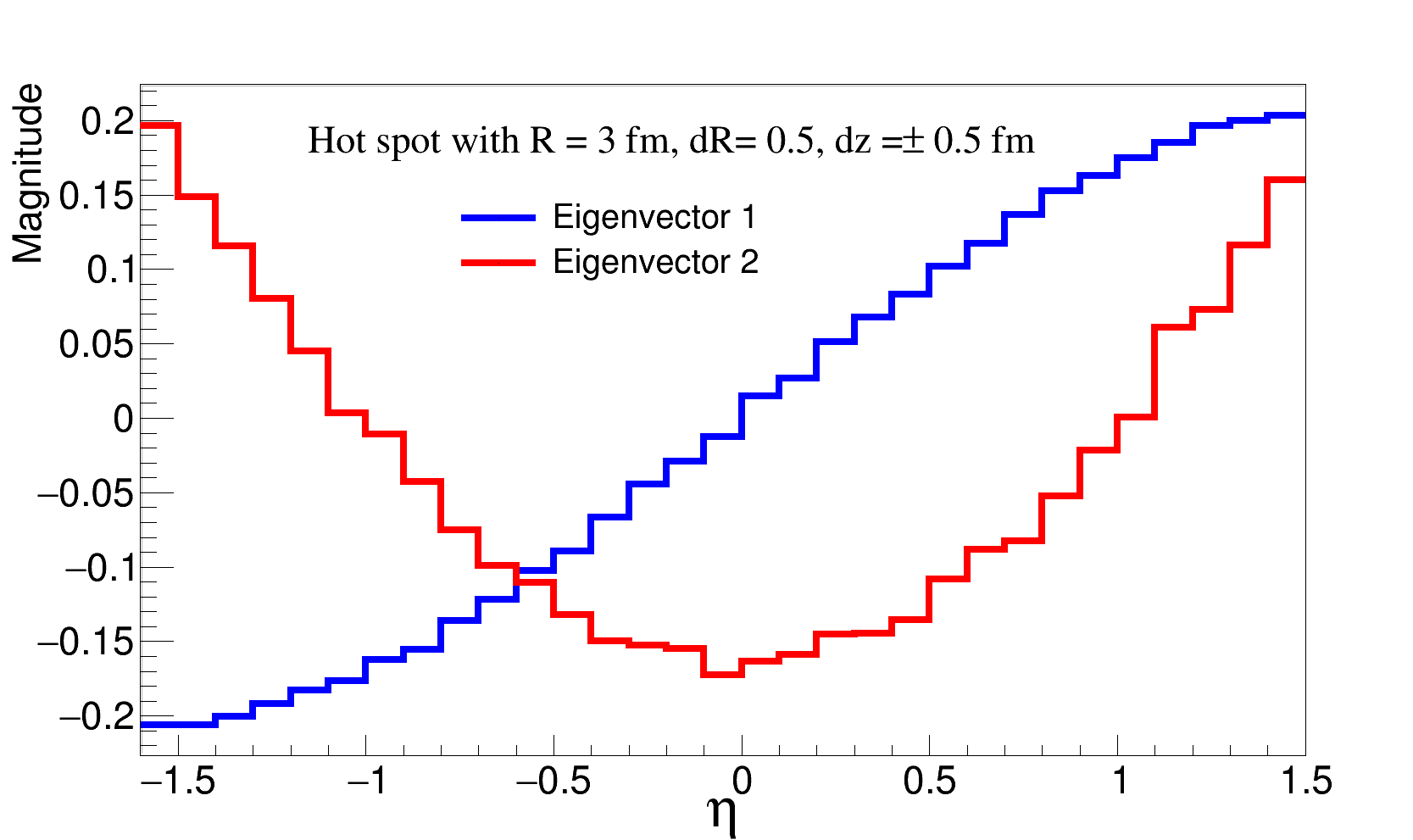}
%	\caption{[Color online] .}
	\caption{\label{eigenvector_eta}The first two eigenvectors obtained from PCA decomposition of $\eta$ distribution for events where hot spots are implemented.}
\end{figure}

\begin{figure}[h!]
	\centering
	\includegraphics[width=85mm,height=5.0cm]{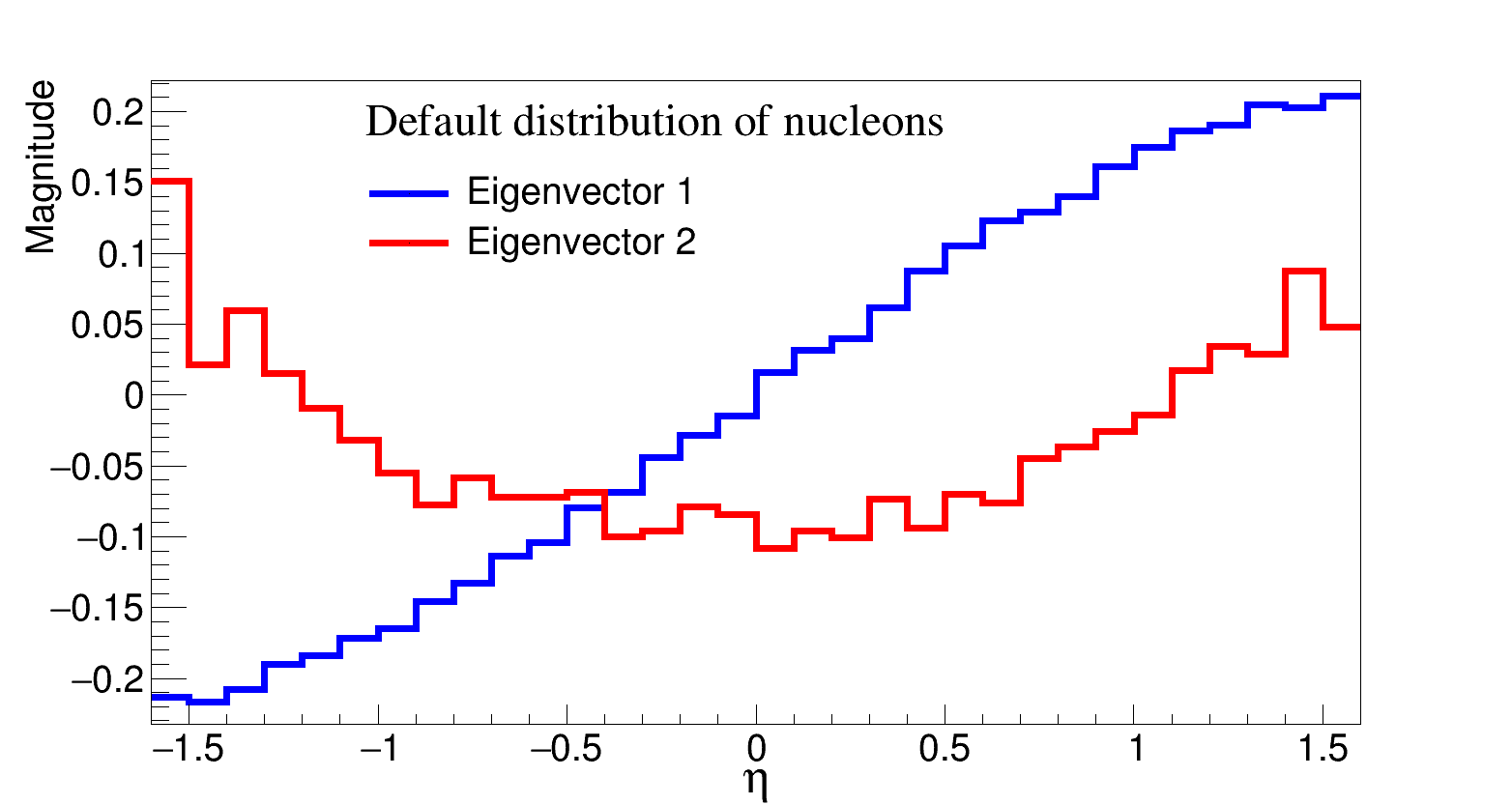}
%	\caption{[Color online] .}
	\caption{\label{eigenvector_eta_default}The first two eigenvectors obtained from PCA decomposition of $\eta$ distribution for events where hot spots are not implemented.}
\end{figure}

\iffalse
It is interesting to note that when we analyze events without introducing hot spots, PC eigenvectors does not show any definite pattern. 
\fi
It may be noted that after hot spots implementation in the initial state the eigenvectors obtained from the PC decomposition of $\phi$-distribution shows more prominent cosine and sine modulation compared to the default nucleonic distribution, as shown in Fig.~\ref{eigenvector_phi_uncluster}.
This may be because that hot spots in the initial state induce more anisotropic flow or flow fluctuations as has been reflected by PCA results. Similarly optimal basis set to represent $\eta$-distribution can also be obtained via its PCA decomposition. Fig~\ref{eigenvector_eta} and ~\ref{eigenvector_eta_default}  shows the first two eigenvectors obtained from PC decomposition of $\eta$ distributions with and without implementation of hot spot in the initial condition, respectively. The decomposition of the $\eta$ distribution in terms of eigenvectors can be expressed as
\begin{equation}
\frac{dN}{d\eta} \approx  c_{1}\eta + c_{2}(A\eta^{2} - B ),
\end{equation} where A and B are arbitrary constant numbers whose exact value may depend on kinematic selections, physics input in the model, etc and c$_{1,2}$ are the coefficients of the eigenvectors.
Use of these basis vectors to extract other observables like flow components, or components representing fluctuations will be investigated in the later studies.

\section{Summary and Discussion}
In high energy heavy ion collisions, the initial nucleonic configurations of the colliding nuclei play a major role in deciding the properties of the created medium. For understanding the evolution of the colliding system through the finally produced particles, it is extremely important to understand the imprints of each of the phases through which the medium evolves. One stage which is least understood so far is the initial state of the nucleons or partons. It has been demonstrated that during the evolution of the medium the initial spatial configurations get transferred to the final state momentum configurations of the produced particles. In this article, we have employed the technique of the Principal Component Analysis on the $\eta$, $\phi$ and p$_{T}$ distributions of the produced particles and examined the eigenvalues of the first principal component which represents the variance of the lower dimension observable. The aim was to study the sensitivity of these eigenvalues to the initial state fluctuations in nucleon positions. We have introduced a spatial rearrangement of nucleonic  configurations at the incoming nuclei to mimic hot spot like energy deposition in UrQMD and compared the PCA results on the distributions of produced pions with and without the implementing the modification. It has been observed that for the analyses of the 1-D distributions i.e., ($\eta$, $\phi$, p$_{T}$) and the 2-D ($\eta$-$\phi$, $\phi$-p$_{T}$ and $\eta$-p$_{T}$) distributions obtained by making combinations of the variables, the eigenvalues increase with the size of the hot spots.
The increase in eigenvalue is more prominent in the azimuthal distributions due to the nature of the hot spot implemented in the transverse plane.
We have also observed that the eigenvalues increase towards the peripheral collisions showing more prominent effect on the azimuthal distributions. For studying a realistic scenario, we have seen that the eigenvalues increase with the increase in the fraction of the events with initial nucleonic hot spots. These observations can be used to filter the events based on PC eigenvalues and then study other observables like the flow coefficients and flow fluctuations in the events with relatively higher eigenvalues. These events are likely to have hot spots in the initial configurations and its effect is likely to be prominently reflected in the final state observables mentioned above. We have also obtained the eigenvectors from the PCA decomposition of $\phi$ and $\eta$ distributions. The 1$^{st}$ and 2$^{nd}$ eigenvectors from the $\phi$-distribution can be identified with sin(2$\phi$) and cos(2$\phi$) Fourier bases, respectively. Similarly, orthonormal basis set for $\eta$ distribution are seen to be associated to a form which appears to be Legendre polynomials but not exactly so. It is however important to understand the sensitivity of the variables to the structures likely to be formed at intermediate stages other than or in addition to the initial stage of the collisions. These investigations are currently beyond the scope of this paper and will be followed up in subsequent studies.

%\section*{Acknowledgement} 
%This research has used resources of grid computing facility at Variable Energy Cyclotron Centre (VECC), Kolkata. 

\bmhead{Acknowledgments}
This research has used resources of grid computing facility at Variable Energy Cyclotron Centre (VECC), Kolkata.

%\section*{References}
%\bibliography{mybibfile}


\bibliography{sn-bibliography}% common bib file


\end{document}